\documentclass[lettersize,journal]{IEEEtran}
\usepackage{amsmath,amsfonts}
\usepackage{textcomp}
\usepackage{stfloats}
\usepackage{url}
\usepackage{verbatim}
\usepackage{graphicx}
\usepackage{cite}
\usepackage{amsmath,amsfonts}
\usepackage{algorithmic}
\usepackage{algorithm}
\usepackage{array}

\usepackage[caption=false,font=footnotesize,labelfont=rm,textfont=rm]{subfig}
\usepackage{bbding}
\usepackage{booktabs}
\usepackage{multirow}
\usepackage{makecell}
\usepackage{amsmath,amsthm,amssymb,amsfonts}
\usepackage{threeparttable}
\usepackage{xcolor} 

\begin{document}
\title{OMLog: Online Log Anomaly Detection for Evolving System with Meta-learning}

\author{Jiyu Tian,~Mingchu Li,~Zumin Wang,~Liming Chen, ~\IEEEmembership{Senior Member,~IEEE},~Jing Qin,~Runfa Zhang

\thanks{This paper is supported by the National Nature Science Foundation of China under Grant Number: T2350710232 and 62466025, the Fundamental Research Funds for the Central Universities under no. DUT20GJ205. (\textit{Corresponding author: Mingchu Li and Zumin Wang})}

\thanks{Jiyu Tian and Mingchu Li are with the School of Software Technology, Dalian University of Technology, Dalian, China. Mingchu Li is also affiliated with the School of Computer and Information Engineering, Jiangxi Normal University, Nanchang, Jiangxi, China. Email: tianjiyu@mail.dlut.edu.cn, mingchul@dlut.edu.cn}

\thanks{Zumin Wang is with the College of Information Engineering, Dalian University, Dalian, China. Email: wangzumin@dlu.edu.cn}

\thanks{Liming Chen is with the School of Computer Science and Technology, Dalian University of Technology, Dalian, China. Email: limingchen0922@dlut.edu.cn}

\thanks{Jing Qin is with the School of Software Engineering, Dalian University, Dalian, China. Email: qinjing@dlu.edu.cn}

\thanks{Runfa Zhang is with the School of Automation and Software Engineering, Shanxi University, Taiyuan, China. Email: zrf@sxu.edu.cn}
}
\markboth{IEEE Transactions on XX}
{How to Use the IEEEtran \LaTeX \ Templates}

\maketitle
\begin{abstract}
Log anomaly detection (LAD) is essential to ensure safe and stable operation of software systems. Although current LAD methods exhibit significant potential in addressing challenges posed by unstable log events and temporal sequence patterns, their limitations in detection efficiency and generalization ability present a formidable challenge when dealing with evolving systems. To construct a real-time and reliable online log anomaly detection model, we propose OMLog, a semi-supervised online meta-learning method, to effectively tackle the distribution shift issue caused by changes in log event types and frequencies. Specifically, we introduce a maximum mean discrepancy-based distribution shift detection method to identify distribution changes in unseen log sequences. Depending on the identified distribution gap, the method can automatically trigger online fine-grained detection or offline fast inference. Furthermore, we design an online learning mechanism based on meta-learning, which can effectively learn the highly repetitive patterns of log sequences in the feature space, thereby enhancing the generalization ability of the model to evolving data. Extensive experiments conducted on two publicly available log datasets, HDFS and BGL, validate the effectiveness of the OMLog approach. When trained using only normal log sequences, the proposed approach achieves the F1-Score of 93.7\% and 64.9\%, respectively, surpassing the performance of the state-of-the-art (SOTA) LAD methods and demonstrating superior detection efficiency. 
\end{abstract}


\section{Introduction}
\label{sec:1}
\IEEEPARstart{S}{oftware} system log files provide a valuable source of information for understanding their behavior and identifying potential anomalies. As scale and complexity increase, anomalous issues such as hardware failures, misconfigurations, and security breaches can lead to performance degradation and system downtime \cite{acmsurvey, identify,  reports, dorm, logformer, parsesurvey, logfit, loggraph}. Log-based anomaly detection (LAD) methods are effective in capturing potential problems and threats and have become powerful tools for automatically monitoring and analyzing system anomaly patterns, including machine learning-based approaches IM \cite{im} and SemPCA \cite{pcatosem}, and deep learning-based approaches DeepLog \cite{deeplog}, LogAnomaly \cite{loganomaly}, LogRobust \cite{logrobust}, PLELog \cite{plelog}, and MetaLog \cite{metalog}.

However, systems may evolve due to function updates, security patches, or infrastructure modifications, posing a significant challenge to existing detection methods. In practice, system evolution may lead to distribution shift as evidenced by changes in event types and event frequencies, including (i) the updating, generation, and deprecation of log event types due to system function changes. For example, when a system adds function modules, new log event types are introduced to record the behavior. Similarly, when the system modifies function modules, the log event types associated with these functions may be replaced by new log event types; (ii) changes in the frequency of log events caused by the external environment. For example, the redesign of business processes may cause log events to occur in a different order or certain event types to occur more frequently.

Currently, offline static detection methods \cite{loganomaly, robustlog, plelog, swisslog, deepsyslog} can fit new log event types by semantic similarity, but cannot address event types generated by completely new functional modules that are semantically unrelated to historical data. Moreover, if the detection model is trained based on fixed distributions, these static methods may not be able to accurately identify anomalies when the sequence pattern changes. Therefore, supervised and semi-supervised dynamic detection methods \cite{ada, roead, logonline} have been proposed and adapted to evolving systems through online learning mechanisms, respectively. However, these methods either need to rely on precious abnormal labels and large-scale storage space \cite{roead}, or the online update strategy cannot satisfy the real-time and generalization requirements \cite{logonline}. We find that existing online LAD methods still face the following challenges:

\begin{itemize}
\item \textbf{Efficiency:} Existing online LAD methods update the model to fit the new sequence pattern before each detection regardless of whether a distribution shift is observed. For example, LogOnline \cite{logonline} needs to utilize a normality detection model to identify high-confidence normal samples among the test samples for online training before each detection; the periodic updating strategy of ADA \cite{ada} reduces the frequency of model updates, but cannot capture new anomalous patterns capture of novel anomalous patterns in a timely. If distribution shifts cannot be effectively recognized, the continuous updates make it difficult to extend the model to large-scale software systems.

\item \textbf{Generalization:} The training strategy of existing online LAD methods makes it difficult to generalize old models to evolving logs. For example, the model training process of the ROEAD \cite{roead} method requires manual feedback on the state of the system and cannot automatically fit the evolving data. LogOnline \cite{logonline} completes the weight update with only one iteration of training using identified high-confidence normal samples. If the limited local information in the evolutionary data cannot be utilized to achieve effective generalization, the model may struggle to maintain high detection performance for changing anomaly patterns.
\end{itemize}

Inspired by the stability of local log event types and the similarity of near-neighbor log samples (described in detail in Section \ref{sec:2.3}), we propose a semi-supervised log anomaly detection method based on online meta-learning, called OMLog, to address the above challenges. The method performs conditional online meta-learning by identifying distribution shifts to enhance the detection efficiency and generalization of the detection model. Specifically, we first introduce a maximum mean discrepancy (MMD)-based distribution shift detection algorithm, which determines whether to perform the online update of the model by the MMD value between the current batch of samples and the historical samples. Furthermore, we develop a semi-supervised online learning mechanism using meta-learning. This mechanism involves creating a meta-task using test data and high-confidence normal sequences identified by a normality detection model, based on the temporal nearest-neighbor principle, then leverages the similarity of nearest-neighbor features to fast generalize the model to local sequence patterns. We evaluated the effectiveness of OMLog on two log datasets, stable HDFS and evolved BGL, respectively, and the results indicate that using only normal log sequences for training, OMLog far outperforms existing SOTA LAD methods in terms of detection performance while being more efficient. In summary, our main contributions are highlighted as follows:

\begin{itemize}
\item We propose OMLog, a semi-supervised online log anomaly detection method for dealing with the problem of log data distribution shift due to system evolution.

\item We introduce a distribution shift detection algorithm based on the stability of local log events, which uses the maximum mean discrepancy value to determine whether to perform online model updating.

\item We design an online meta-learning method based on the similarity of near-neighbor log samples, which combines high-confidence normal samples with test samples according to temporal nearest-neighbors into a meta-training task for online updating to enable fast generalization to the evolving log.

\item We conduct extensive experiments on public log datasets, and the experimental results demonstrate the detection efficiency and performance of OMLog.
\end{itemize}

\section{Background and Motivation}
\label{sec:2}
\subsection{Log Terminology}
\label{sec:2.1}

\begin{figure}[h]
  \centering
  \includegraphics[width=\linewidth]{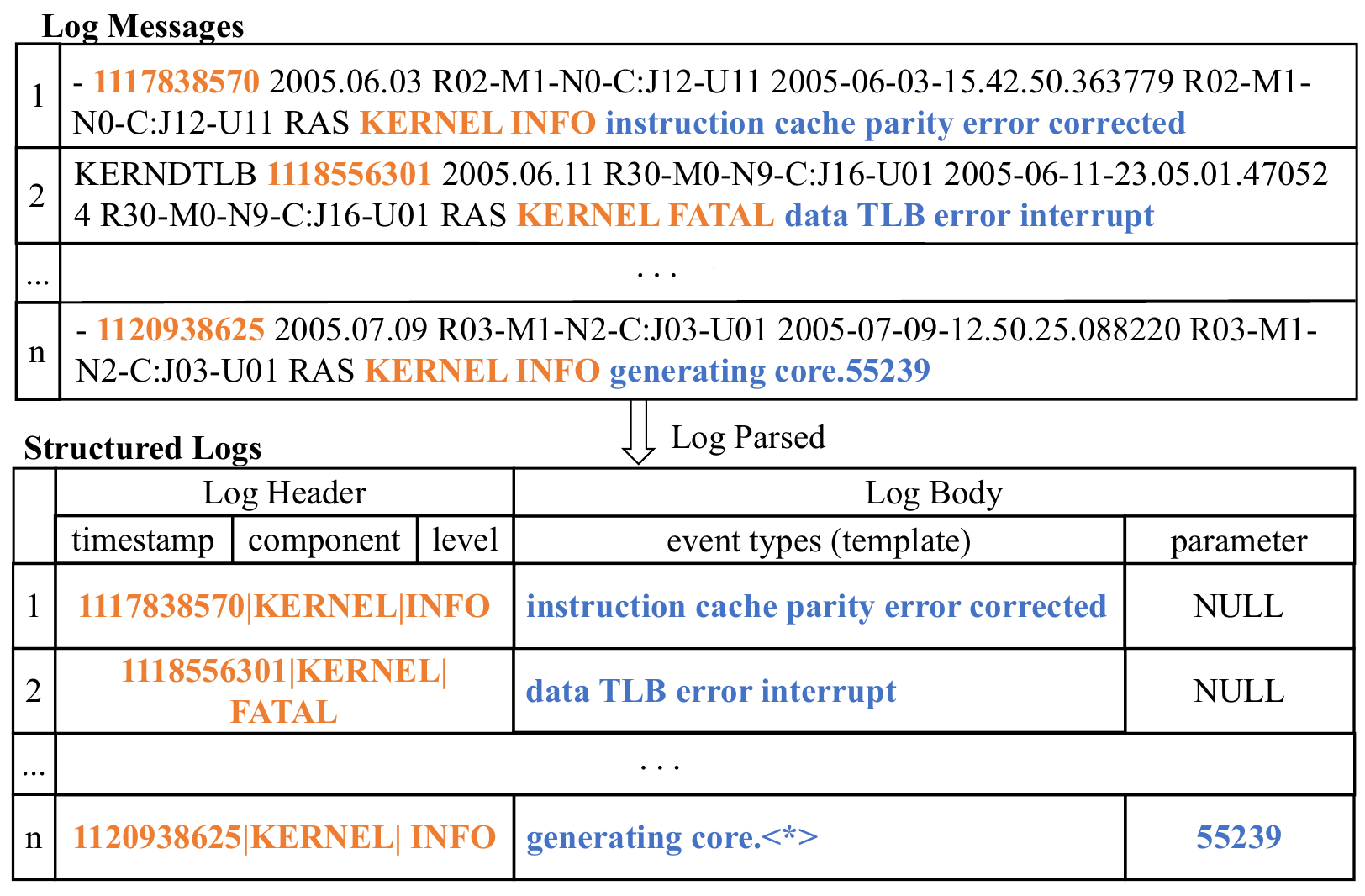}
  \caption{An illustrative example of log terminology based on BGL dataset.}
  \label{figure1}
\end{figure}

In complex software systems, logs are valuable resources for recording the operational status of the system \cite{loggram}. As shown in Figure \ref{figure1}, logs are recorded in text form and usually consist of two parts: log header and log body. The log header provides contextual information about the events and is mainly composed of timestamps, components, and levels. The log bodies are the core content in the log, they can be considered as the output of print statements in the source code, including log events and parameters describing the system state.

The process of separating log bodies into log events and parameters is known as log parsing. Log parsing~\cite{loggram} is often regarded as an upstream task for log anomaly detection, which can convert unstructured log messages into structured log data for subsequent analysis and detection. For example, as shown in Figure \ref{figure1}, the original log message "generating core.55239" is parsed into the log event "generating core.\textless*\textgreater" and the parameter "55239", where the parameter is represented by the placeholder \textless*\textgreater in the log event. We arrange the log events in chronological order to form a log sequence and then group the log sequence according to a session, fixed or sliding window to obtain log samples. The log samples can completely reflect the activities of the system during a certain period and are the data basis for offline and online log anomaly detection.

\begin{figure*}[!htb]
	\centering
	\scriptsize
	\subfloat[Normal Base-Part1]{
		\includegraphics[width=0.155\linewidth]{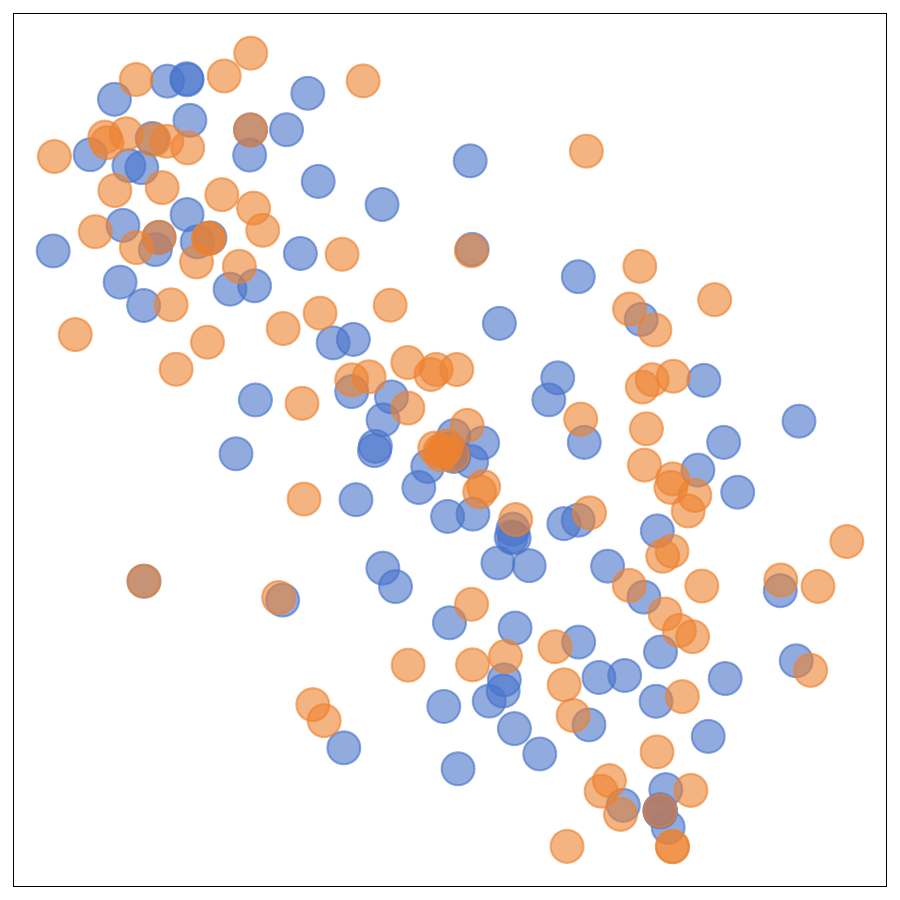}
	}
	\hfill
	\subfloat[Normal Base-Part2]{
		\includegraphics[width=0.155\linewidth]{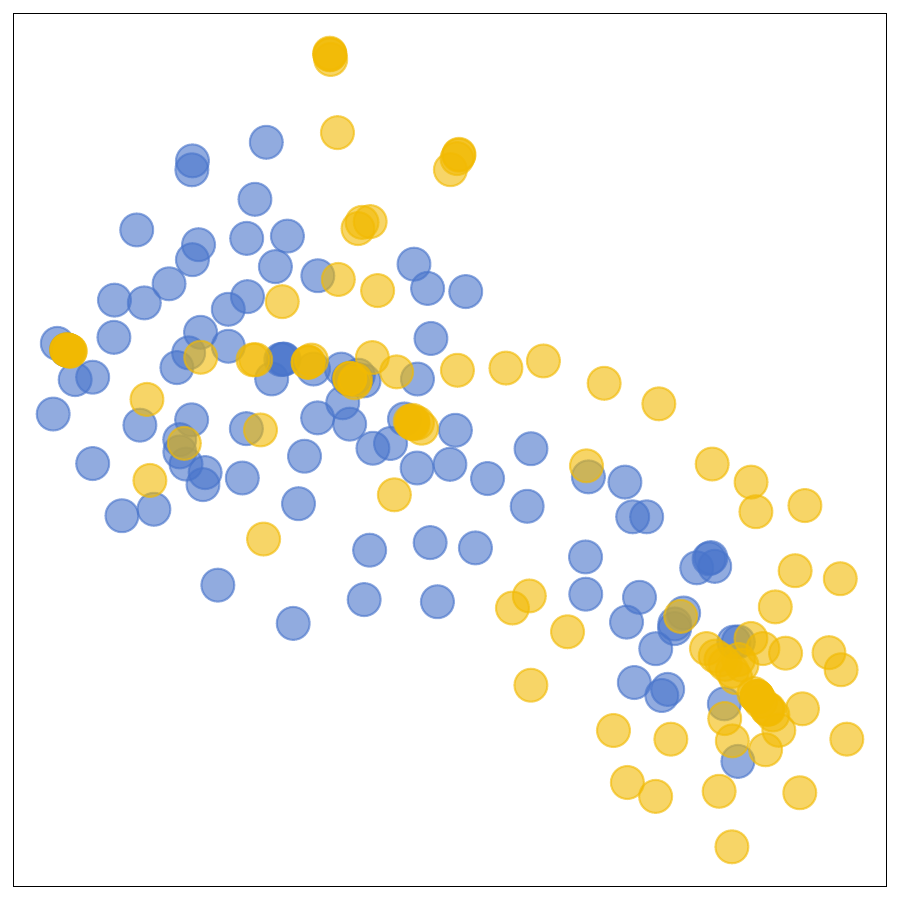}
	}
	\hfill
	\subfloat[Normal Base-Part3]{
		\includegraphics[width=0.155\linewidth]{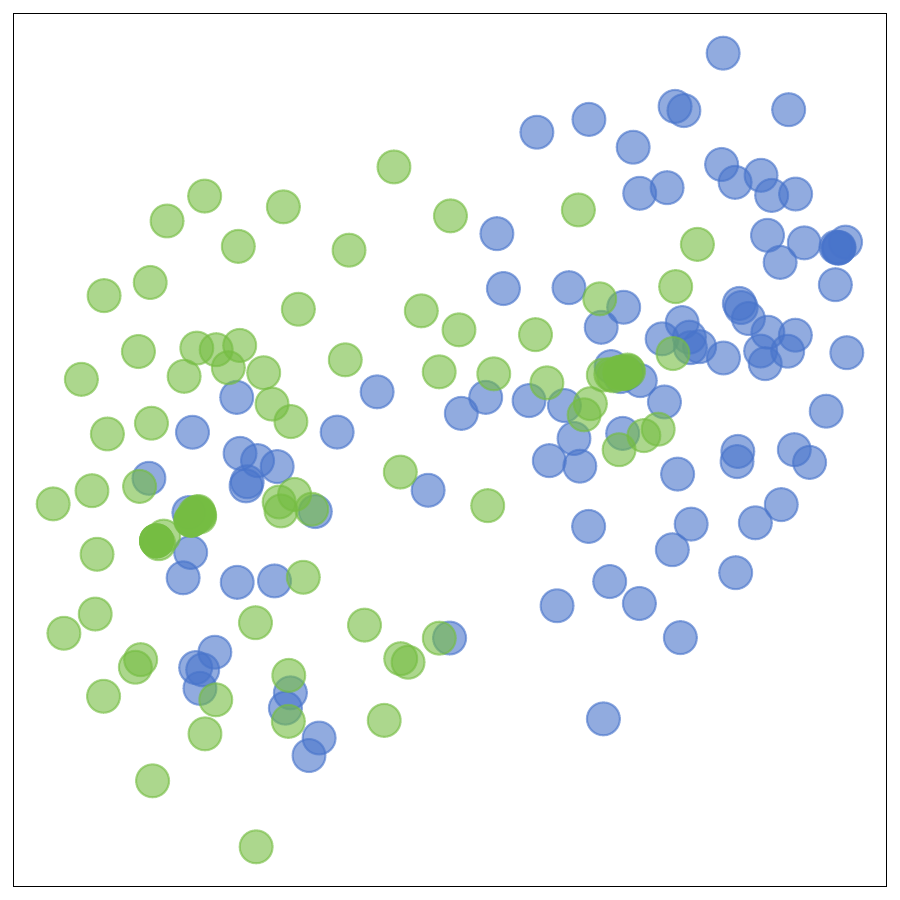}
	}
	\hfill
	\subfloat[Normal Base-Part4]{
		\includegraphics[width=0.155\linewidth]{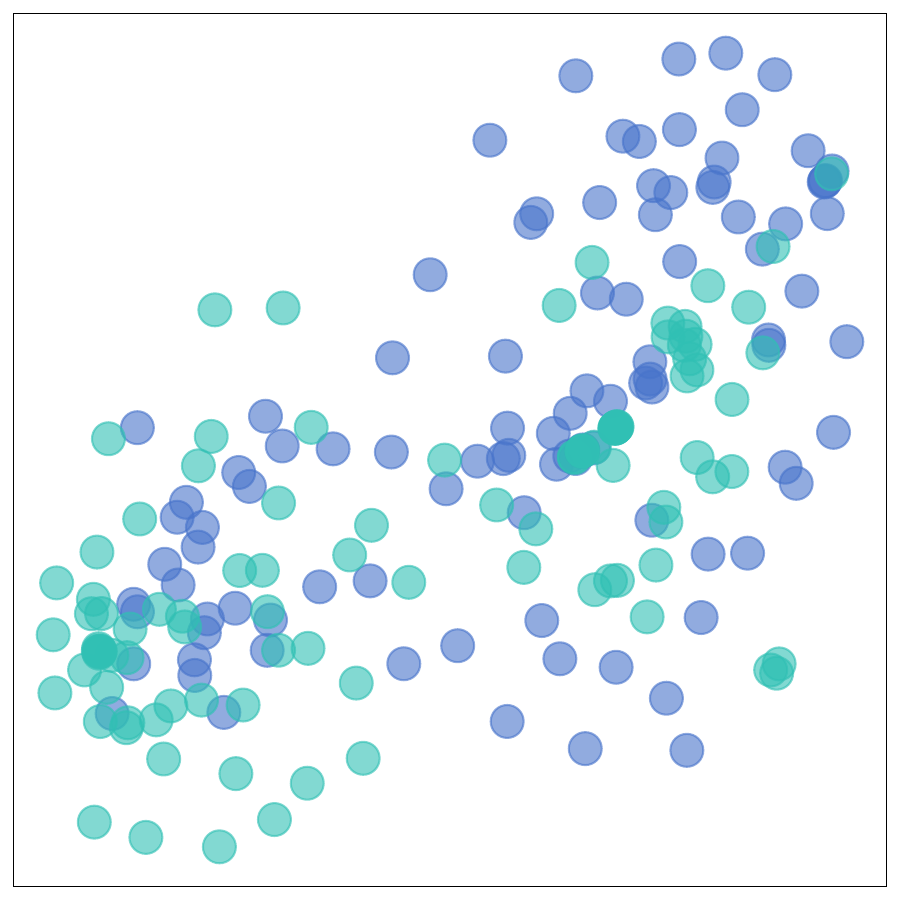}
	}
 	\hfill
	\subfloat[Normal Base-Part5]{
		\includegraphics[width=0.155\linewidth]{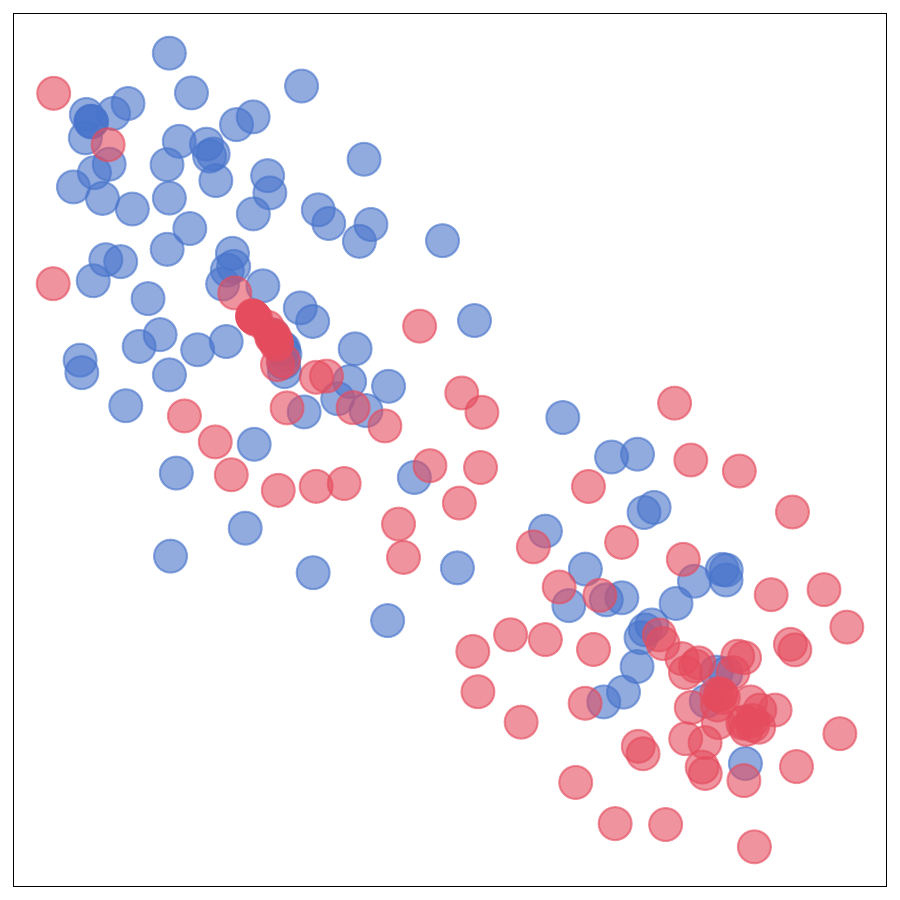}
	}
	\hfill
	\subfloat[Normal Base-Part6]{
		\includegraphics[width=0.155\linewidth]{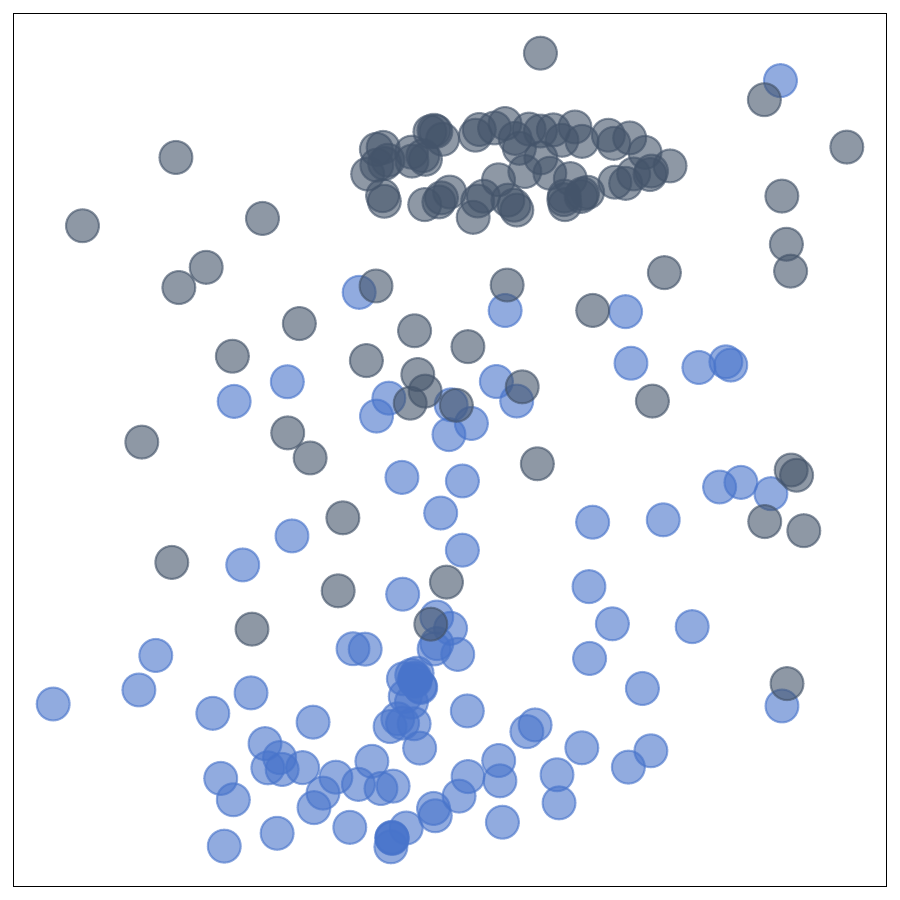}
	}

	\centering
	\scriptsize
	\subfloat[Abnormal Base-Part1]{
		\includegraphics[width=0.155\linewidth]{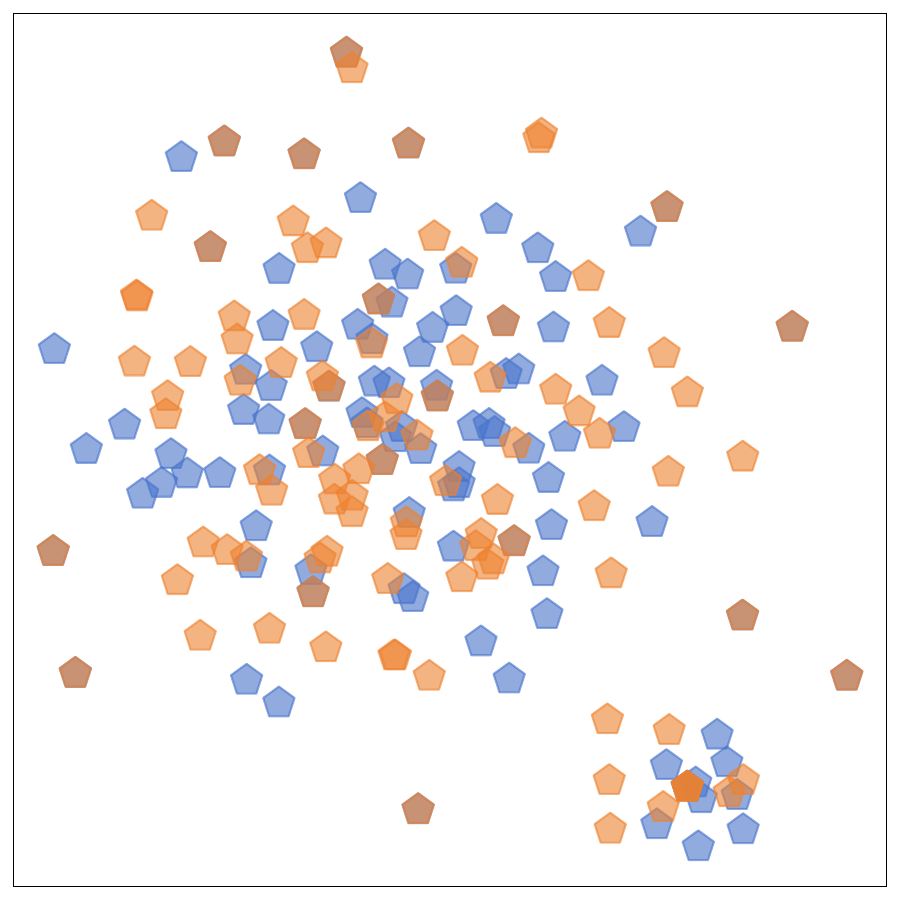}
	}
	\hfill
	\subfloat[Abnormal Base-Part2]{
		\includegraphics[width=0.155\linewidth]{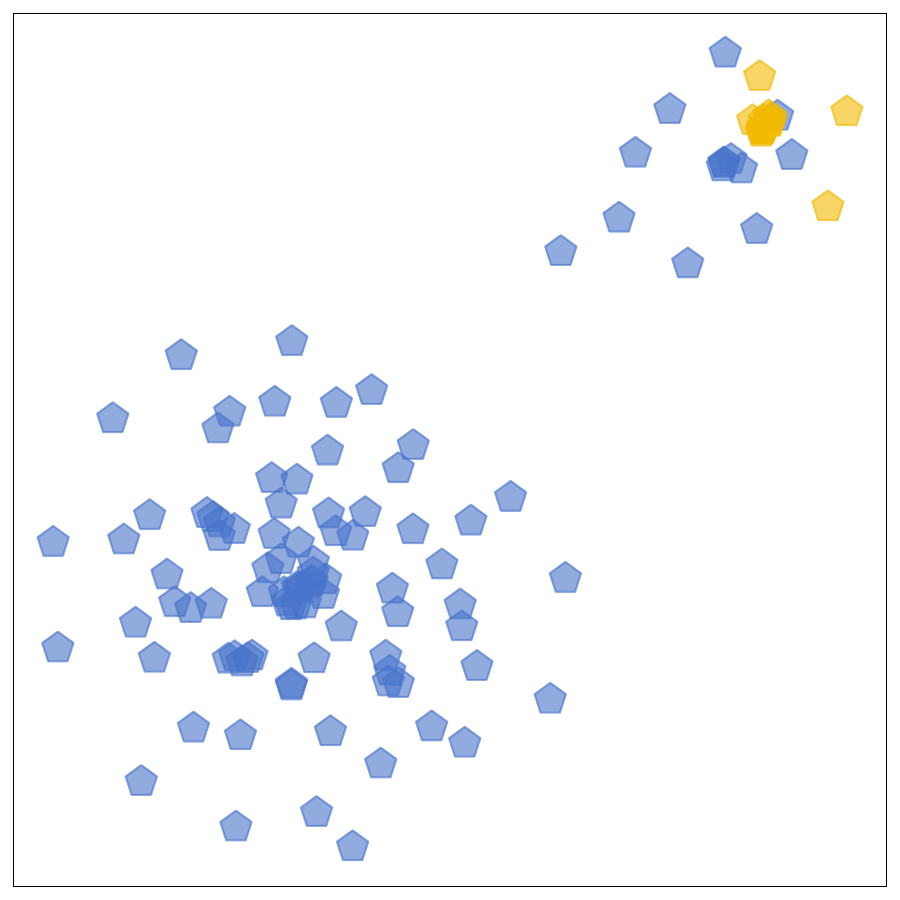}
	}
	\hfill
	\subfloat[Abnormal Base-Part3]{
		\includegraphics[width=0.155\linewidth]{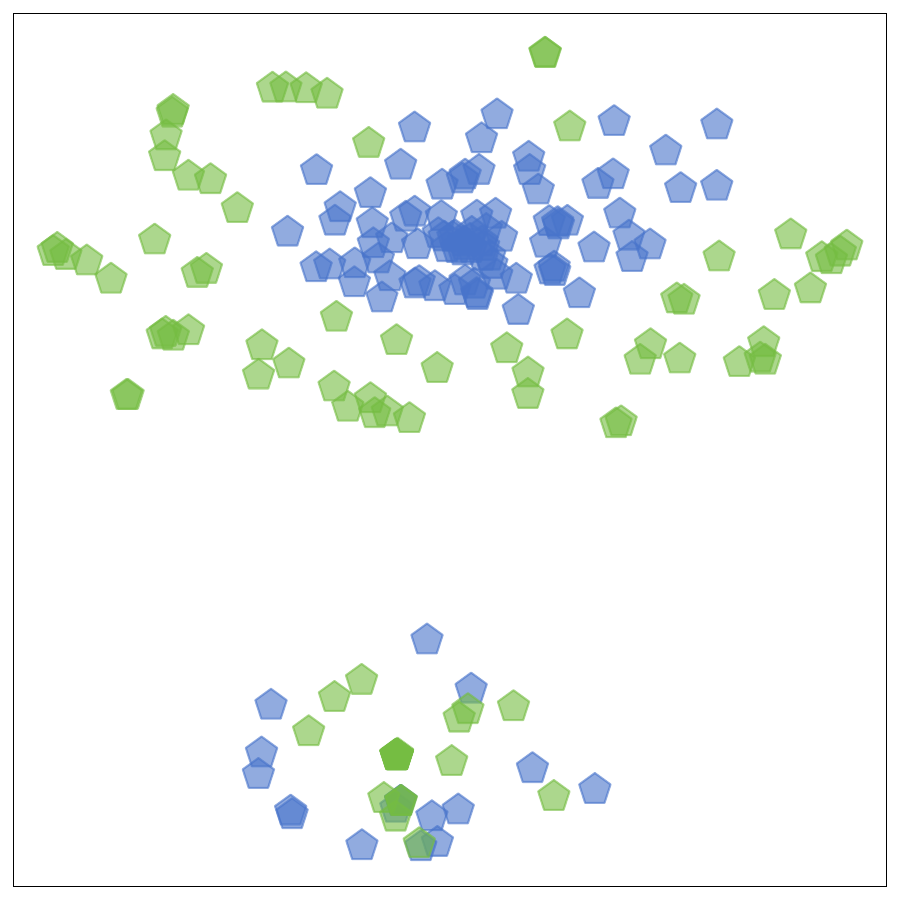}
	}
	\hfill
	\subfloat[Abnormal Base-Part4]{
		\includegraphics[width=0.155\linewidth]{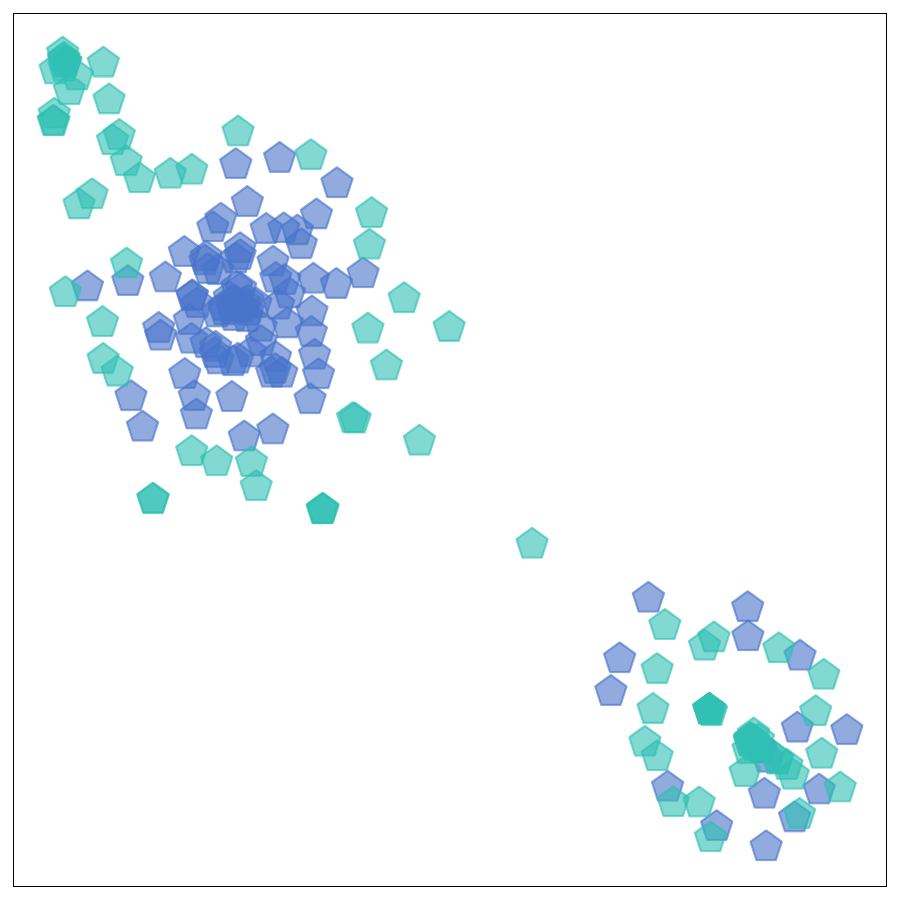}
	}
 	\hfill
	\subfloat[Abnormal Base-Part5]{
		\includegraphics[width=0.155\linewidth]{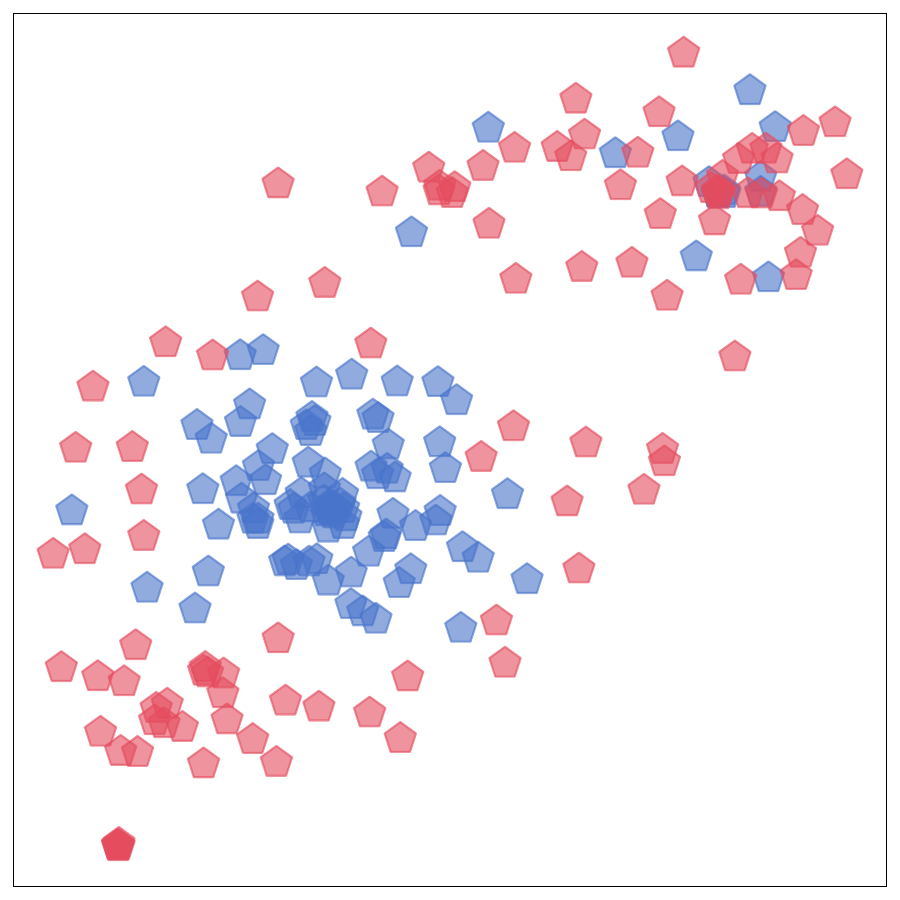}
	}
	\hfill
	\subfloat[Abnormal Base-Part6]{
		\includegraphics[width=0.155\linewidth]{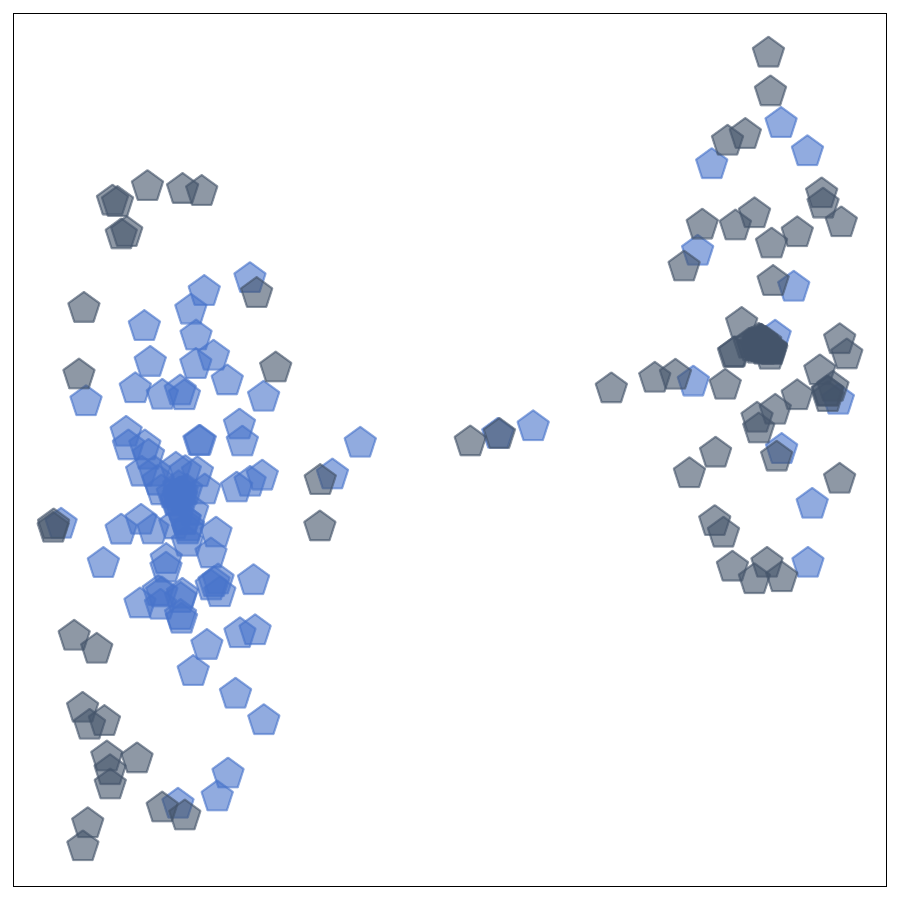}
	}
 
        \centering
        \scriptsize
	\subfloat{
		\includegraphics[width=0.7\linewidth]{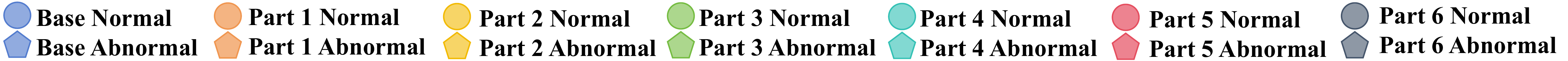}
	}
\caption{Intuitive demonstration of distribution shifting from BGL log data. The datasets are constructed utilizing the sliding window with a window size of 100, a step size of 100, and is divided into 6 parts in chronological order. Initially, we randomly selected 200 samples from Part 1 to serve as the base part for comparison. Subsequently, we randomly selected 200 samples from each Part, where each subset comprised 100 normal samples and 100 abnormal samples. Subfigures (a)-(f) show examples of normal samples and (g)-(l) show examples of abnormal samples.}
\label{figure2}
\end{figure*}

\subsection{Online Log-based Anomaly Detection}
\label{sec:2.2}

System evolution has rendered the static log anomaly detection model incapable of addressing new event types and abnormal patterns. Therefore, online log anomaly detection methods have emerged, which can dynamically update the model while detecting the data to adapt to unstable log data. Currently, the widespread online methods mainly include ROEAD \cite{roead} and LogOnline \cite{logonline}. ROEAD is a supervised log-based anomaly detection method, which aims to improve support vector machines to enhance the detection efficiency by utilizing the online mirror descent algorithm. However, the limitation of labels makes ROEAD difficult to apply in practice. LogOnline based on semi-supervised learning utilizes only normal samples to update model parameters, getting rid of the dependence on labels. Its core components mainly include next event prediction and normality detection model.

\textbf{Next Event Prediction:} The system runs according to a fixed process when no anomaly occurs. Therefore, the next event in the log sequence is predictable, and log events outside the prediction range can be considered as anomalies. Since the log sample is a time sequence of different log event types. Therefore, LogOnline follows the work of \cite{deeplog, loganomaly} and employs a long short-term memory (LSTM) \cite{deeplog,loganomaly,logrobust} neural network to build a multi-classification model. The output of the model is the conditional probability distribution of the samples corresponding to each log event. For example, in the training phase, a log sequence is assumed to have 6 log events: \{e1, e2, e3, e4, e5, e6\}. When the window size h = 3, the inputs and outputs corresponding to the next event prediction models are: \{e1, e2, e3→e4\}, \{e2, e3,e4 →e5\}, \{e3, e4, e5→e6\}. In the evaluation phase, the predicted probability of each sample is first ranked according to the model. The next log event is considered normal if it is observed within the Top-K candidate events predicted by the model. Noting that the evolution of the system leads to new events, LogOnline proposes a variable classifier to automate the adjustment of the model's output categories

\textbf{Normality Detection Model:} Detection methods based on next-event prediction require a continuous reliance on normal sequences to update the model online. Therefore, LogOnline provides a normality detection model that utilizes additional log header information to determine the confidence level of normal samples in unseen test data. In practice, the log header of each log contains a timestamp, component name, and redundancy level information. The normality detection model uses the log header information in the normal instances to train an auto-encoder model that updates by minimizing the reconstruction error measured by MSE Loss. The model sets log event windows with MSE errors less than a predefined threshold to be considered normal and uses these normal log windows for online learning of the anomaly detection model. It is worth noting that LogOnline needs to go through three processes: normality model detection, online model updating, and anomaly model detection, regardless of whether the distribution of the new data has changed. Therefore, it is difficult for this method to achieve real-time response to large-scale data. Meanwhile, we find that the inference speed of offline detection for a batch of data is extremely fast, and the bottleneck of the LogOnline detection mechanism is online weight update.
\subsection{Distribution Shift, Local Log Event Stability, and Near-neighbor Log Sample Similarity in Evolving Systems}
\label{sec:2.3}

\textbf{Distribution Shift:} As mentioned above, unstable logs during system evolution may lead to significant changes in sequence patterns, which in turn lead to distribution shifts between different batches of log samples from the same system. As shown in Figure \ref{figure2}, in the BGL dataset, the normal and abnormal samples are divided into six parts in chronological order and visualized by the T-SNE algorithm. For the normal samples, it is easy to find that the base and part 1 conform to the assumption of the independent and identically distribution (IID), but show a significant distribution gap with the subsequent data (especially the last part of the normal data). For the abnormal samples, the distribution shift of the input features is more significant. The part 1 exhibits clear boundaries with all the other five parts of the data. Given the above analysis, we clearly define the instability problem of logs in evolving systems as a distribution shift problem. 

\begin{figure}[!htb]
	\centering
	\scriptsize
	\subfloat[Batch Size=100]{
		\includegraphics[width=0.98\linewidth]{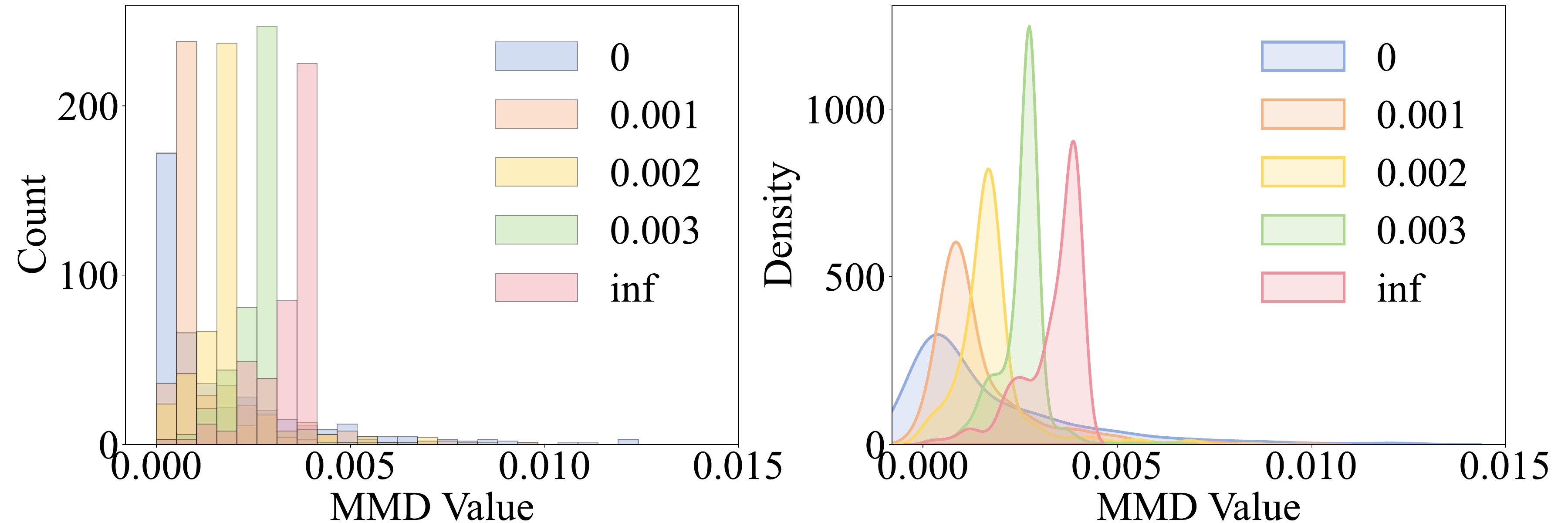}
	}
 
	\hfill
	\subfloat[Batch Size=200]{
		\includegraphics[width=0.98\linewidth]{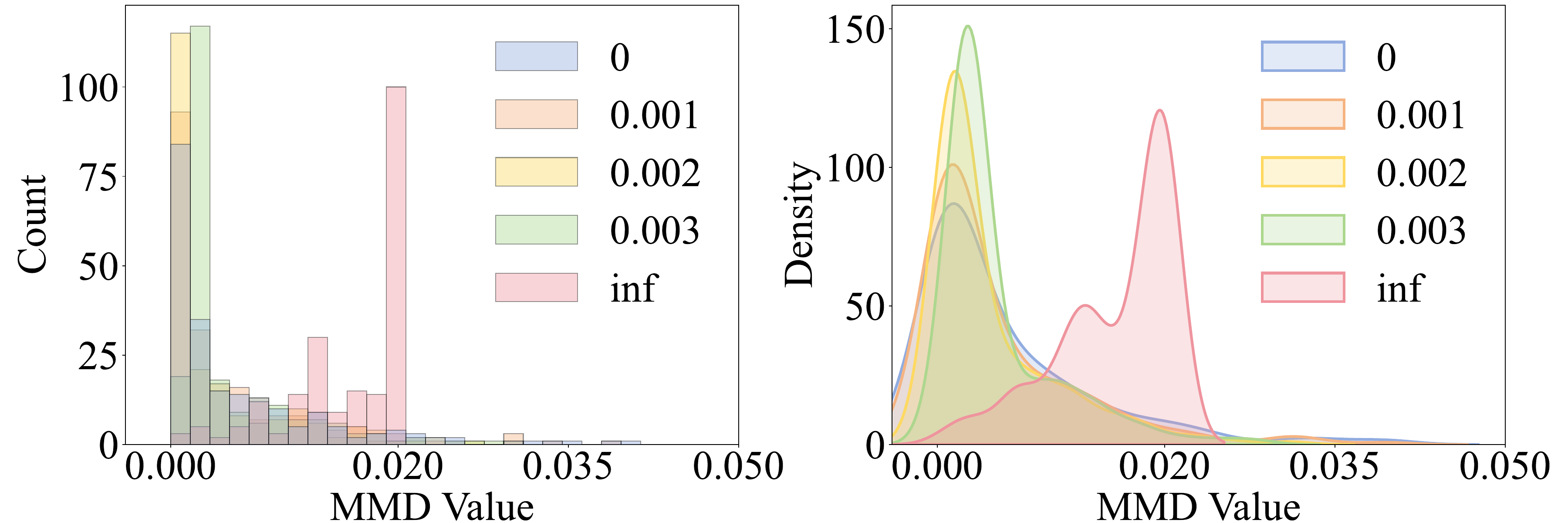}
	}

\caption{Distribution shift between batches of data in the BGL dataset.}
\label{figure3}
\end{figure}

\textbf{Local Log Event Stability Analysis:} In the real world, the function modules and external environment of a system do not change all the time. Most of the time, the system maintains a stable state where the event types and event frequencies remain unchanged, which allows existing detection models to be directly applied to the next batch of test samples. Taking the BGL dataset as an example, if set 100/200 samples as a batch, a total of 425/200 batches can be obtained. Among them, 42 batches have the same data as the previous batch. As shown in Figure \ref{figure3}, if we set the MMD threshold for distribution shift to 0.001, there are 264 batches with no distribution shift compared to the previous batch of data, accounting for 62.1\% of the total batches (50\% with 200 samples as a Batch). It can be demonstrated that the system evolution only occurs at certain times. If we conduct online detection on evolved log data and perform rapid offline inference on stable log data, we can conserve substantial computational resources.

\begin{figure}[!htb]
	\centering
	\scriptsize
	\subfloat[Batch Size=100]{
		\includegraphics[width=0.98\linewidth]{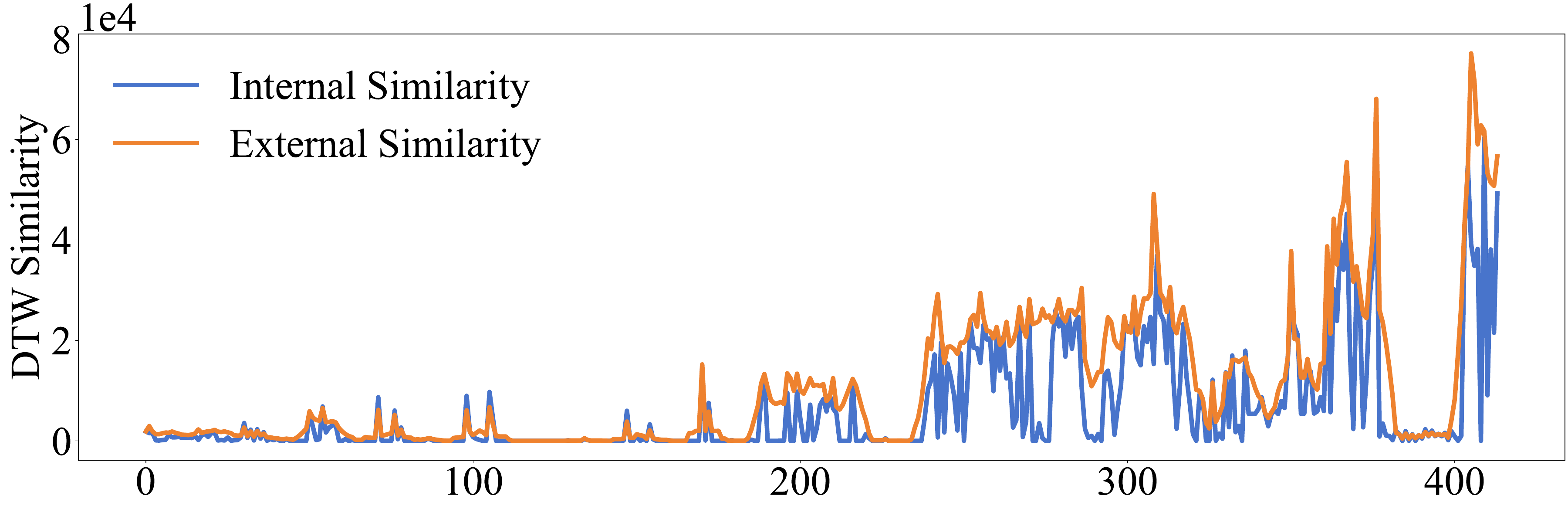}
	}
 
	\hfill
	\subfloat[Batch Size=200]{
		\includegraphics[width=0.98\linewidth]{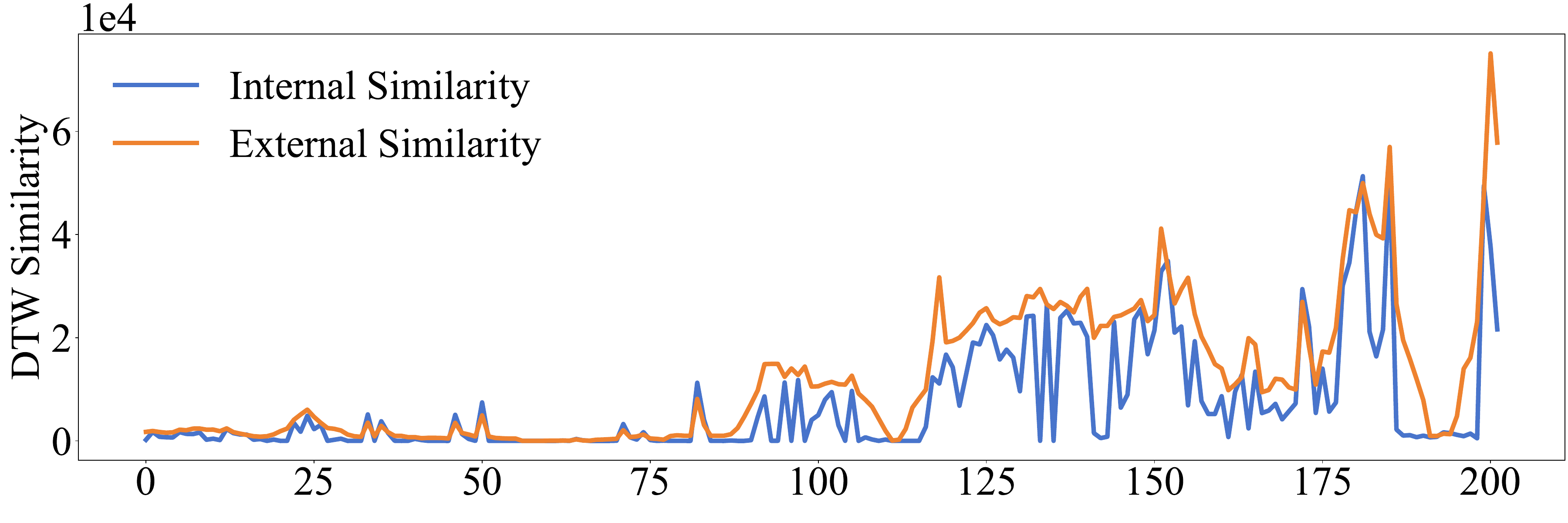}
	}

\caption{Sample similarity of the BGL dataset based on dynamic Time Warping. Internal similarity, i.e., the average of the similarity between each sample within a batch of data; External similarity, i.e., the similarity between each sample of the current batch of data and each sample of the previous 10 batches of data.}
\label{figure4}
\end{figure}

\textbf{Near-neighbor Log Sample similarity Analysis:} The periodic activities of a system may generate a large number of repetitive log event types within a period. Thus, the feature spaces of the log samples consisting of these events are highly overlapping and exhibit near-neighbor sample similarity. Taking the BGL dataset as an example, as mentioned above, if consecutive 100/200 log samples are composed into one batch of log data, 425/212 batches can be obtained. We utilize the widely used dynamic time warping (DTW) \cite{dtwnet} algorithm to calculate the similarity between two log sequence samples. As shown in Figure \ref{figure4}, we take the similarity between the samples within the batch as the internal similarity, and the average similarity between the sample of the current batch and the sample of the previous 10 batches as the external similarity, and the resulting comparison between the internal similarity and the external similarity is shown in Figure \ref{figure4}. It can be observed that the similarity of the closer internal is much higher than the similarity of the more distant external. It is easy to understand that when the training samples are very similar to the test samples, the model can usually learn and generalize to the test data more easily.

\section{Methodology}
\label{sec:3}

\subsection{Overview}
\label{sec:3.1}
\begin{figure*}
\centering
\includegraphics[width=0.98\linewidth]{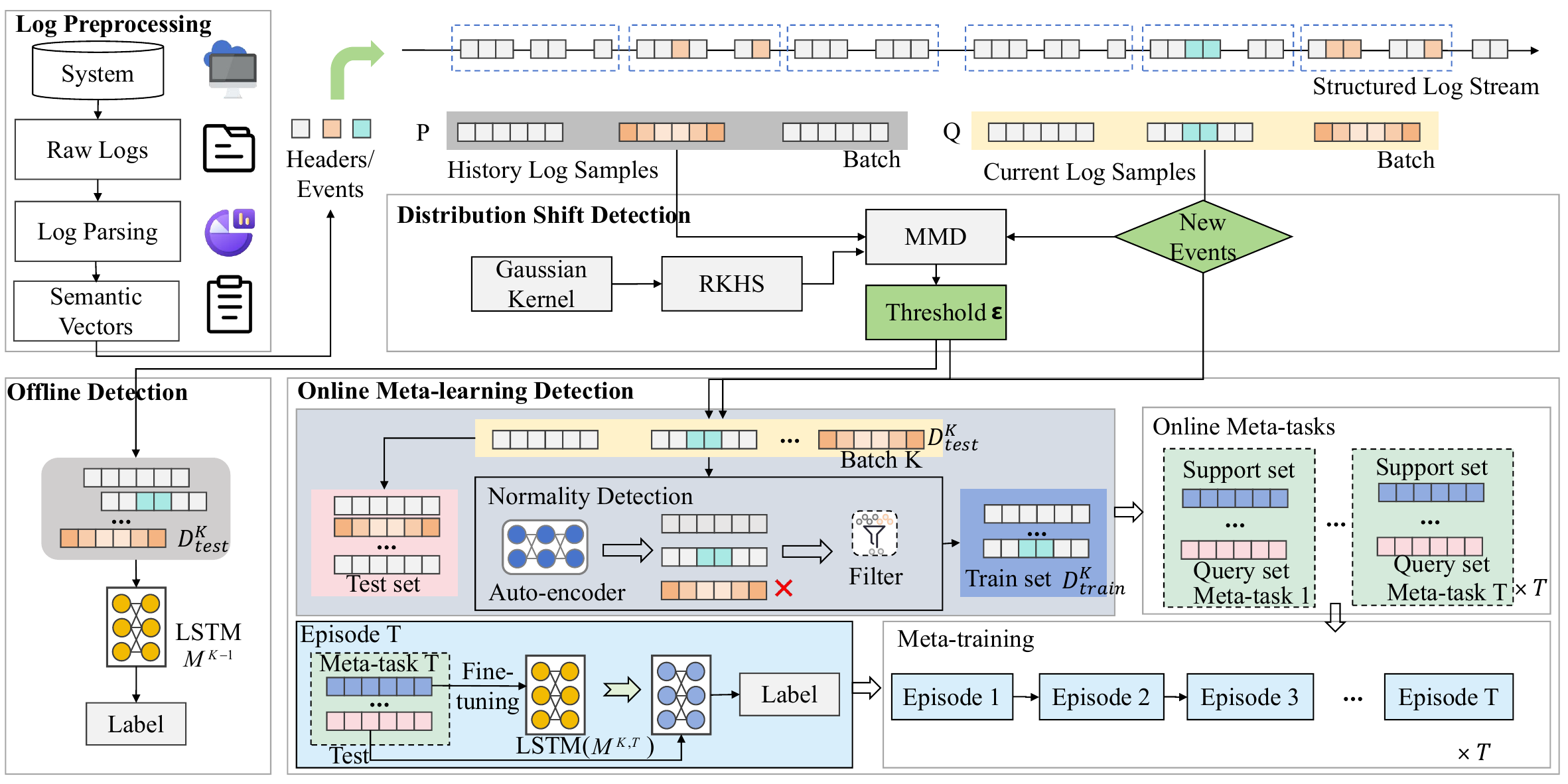}
\caption{Overview of OMLog approach. OMLog implements semi-supervised learning based on next-event prediction, with an auto-encoder for the normality detection model and an LSTM for the anomaly detection model. The novel components of OMLog are DSD and OMD. } 
\label{framework}
\end{figure*}

In this paper, we propose the more practical OMLog, a semi-supervised log anomaly detection method based on online meta-learning, which utilizes easily accessible normal log sequences to handle the log anomaly detection task for system evolution. Specifically, we first obtain sequence samples consisting of structured log headers and log events through log pre-processing, and then use distribution shift detection to determine whether the next batch of log data has evolved significantly. If log evolution occurs, online meta-learning is utilized to fit the gap between the old and new data, otherwise offline detection is utilized to achieve fast inference. As shown in Figure \ref{framework}, OMLog consists of four main components: log preprocessing, distribution shift detection (DSD), online-meta detection (OMD), and offline detection. 

\textbf{Log Preprocessing:} OMLog uses the classical log parsing method Drain \cite{drain} to process unstructured raw logs from the system to obtain log headers (including timestamps, components, and levels) and log bodies. The structured log headers and event types form log sequences in chronological order. We then follow the semantic embedding approach from previous work \cite{logonline} to provide semantic similarity information. Finally, we group the log sequences using session windows and sliding windows to obtain log samples and complete the preprocessing of the raw log data.

\textbf{Distribution Shift Detection (DSD)} utilizes MMD to calculate the difference value between two log sample distributions as an indicator of whether a distribution shift has occurred. First, DSD determines whether the test data contains log event types not seen in the historical data. If the test data contains new event types, online meta detection is triggered directly. Second, inspired by the local log events stability, DSD calculates the difference value between the historical data distribution $\mathcal{P}$ and the test data distribution $\mathcal{Q}$ as the MMD value. If the MMD value is higher than a predefined threshold, the new test samples are considered to have a significant distribution shift from the historical data, triggering online meta Detection. Otherwise, the distribution shift between $\mathcal{P}$ and $\mathcal{Q}$ is considered to be non-existent, and fast inference is achieved by offline detection (more details in Section \ref{sec:3.2}).

\textbf{Online Meta-learning Detection (OMD)} is used to achieve optimal local detection performance by effectively generalizing to the test data after detecting the distribution shift. First, OMD filters the high-confidence normal log sequences from a batch of test data $D_{test}$ through the normality detection model as the training samples $D_{train}$ for online updating. Second, due to the similarity of near-neighbor log samples, OMD selects the samples from the $D_{train}$ as the support set $D_{S}$ to be combined with the query set $D_{Q}$ from the test data to form a meta-task and performs episode training and inference according to the meta-learning approach. (more details in Section \ref{sec:3.3}).

\textbf{Offline detection} is achieved by leveraging the detection model obtained from the previous task. This approach enables rapid inference for fast detection, provided that the DSD component has determined the absence of a significant distributional shift between the data in the current task batch and the data in the previous task batch. Given the similarity in data distribution, the model from the previous task possesses sufficient generalization capability to be applied to the current task effectively.

\subsection{Distribution Shift Detection}
\label{sec:3.2}

The online testing mechanism, which requires updating the weights while testing, can effectively fit distribution gaps, but is computationally intensive and time-consuming. Inspired by the stability of local log events, we determined that realistic systems remain stable most of the time and are only evolving for a small portion of the time when the system is updated or the external environment changes. This implies that not every log data is an evolved log and requires online training to fit it.

Therefore, we enhance the existing approach in two ways: i) using a batch of test data as a test task instead of batch size = 1 in the standard online learning mechanism; ii) designing a distribution shift detection component to analyze the test data for gaps between the test data and the previous batch of training data, and online detecting the evolving data only.

\textbf{MMD for Online Distribution Shift Detection:} MMD can determine the similarity of two distributions by calculating the difference between their samples. Compared to the traditional distance metric \cite{distance}, MMD does not require any assumptions about the form of the distribution of the data, which makes it very suitable for complex and distribution-unknown scenarios. Moreover, the computational complexity of the MMD algorithm depends on the number of samples, which allows it to handle long sequential log data effectively. 

For the online detection of log samples, we define two distributions $\mathcal{P}$, $\mathcal{Q}$, and a set of mapping functions $\mathcal{F}$. Where the log samples $S^{\mathcal{P}}$ in $\mathcal{P}$ consist of online training data from the previous batch, $\mathcal{P}$ = \{$S_1^{\mathcal{P}}$, $S_2^{\mathcal{P}}$, $S_3^{\mathcal{P}}$, ... , $S_p^{\mathcal{P}}$\}, the log samples $S^{\mathcal{Q}}$ in $\mathcal{Q}$ is consist of the current test data, $\mathcal{Q}$ = \{$S_1^{\mathcal{Q}}$, $S_2^{\mathcal{Q}}$, $S_3^{\mathcal{Q}}$, ... , $S_q^{\mathcal{Q}}$\}; the mapping function $f\in\mathcal{F}$, then the MMD value between the training data $\mathcal{P}$ and the test set $\mathcal{Q}$ is:
\begin{align}
\label{E1}
\mathrm{MMD}[\mathcal{F},p,q]=\sup_{f\in\mathcal{F}}(\mathbf{E}_{\chi\sim p}[f(\mathcal{P})]-\mathbb{E}_{\mu\sim q}[f(\mathcal{Q})])
\end{align}

When the number of samples is limited, it can be converted:
\begin{align}
\label{E2}
\mathrm{MMD}\left[\mathcal{F},p,q\right]=\sup_{f\in\mathcal{F}}\left(\frac{1}{p}\sum_{i=1}^{p}f(\mathcal{P}_{i})-\frac{1}{q}\sum_{j=1}^{q}f(\mathcal{Q}_{j})\right)
\end{align}

It is easy to find that the mapping function $\mathcal{F}$ of Eq. (\ref{E2}) maps all the samples in $\mathcal{P}$ and $\mathcal{Q}$ to the corresponding function values, and then a maximum mean discrepancy can be obtained by calculating the mean differences between them and comparing them. Since the computation of MMD relies on the set of mapping functions $\mathcal{F}$, the $f$ should be rich enough but restricted in number to facilitate the computation. Thus, the standard MMD chooses the Gaussian Kernel function as the mapping function to map the raw data to the unit sphere space in the Regenerative Kernel Hilbert Space (RKHS) to fulfill the above requirement, i.e., the Gaussian Kernel function is used to compute the mean of the log samples of $\mathcal{P}$ and $\mathcal{Q}$ in the RKHS. The formula for MMD Value is as follows:
\begin{align}
\label{E3}
\begin{split}
MMD~Value &= \mathrm{MMD}^{2}[\mathcal{F},p,q] \\
&=\|\frac{1}{p}\sum_{i=1}^{p}k\left(S_i^{\mathcal{P}},\cdot\right)-\frac{1}{n}\sum_{j=1}^{q}k\left(S_j^{\mathcal{Q}},\cdot\right)\|^{2} \\
&=\frac{1}{p^{2}}\sum_{i,j=1}^{p}k(S_i^{\mathcal{P}},S_j^{\mathcal{P}})-\frac{2}{pq}\sum_{i,j=1}^{p,q}k(S_i^{\mathcal{P}},S_j^{\mathcal{Q}}) \\
&+\frac{1}{q^{2}}\sum_{i,j=1}^{q}k(S_i^{\mathcal{Q}},S_j^{\mathcal{Q}})
\end{split}
\end{align}
where $k(u,v)$ is the Gaussian kernel function, $k(u,v)=e^{\frac{-||u-v||^{2}}{\sigma}}$

\textbf{Threshold and Judgment:}
After obtaining the MMD value, we judge whether the two distributions are the same or not according to a predefined threshold $\epsilon$. If the MMD value exceeds the threshold $\epsilon$, the two distributions are considered to be significantly different, and the online detection is triggered to generalize the new data using refined meta-learning. Otherwise, the current batch of data is considered not different from the previous batch, triggering offline detection for fast inference. It should be noted that the threshold $\epsilon$ must be set strictly to ensure that an effective fit is achieved for data with significant gaps. If the threshold is too high, test samples that differ from the training data are missed, preventing the old detection model from capturing new anomalous patterns. If the threshold $\epsilon$ is too low, it puts the model in frequent online weight updates and cannot handle large-scale log data. In our implementation, the threshold $\epsilon$ for the distribution shift is empirically set to 1/10 of the average MMD value between different batches of training data.

\subsection{Online Meta-learning Log Anomaly Detection}
\label{sec:3.3}

Online log anomaly detection methods need to address a continuous stream of test tasks. Each test task K corresponds to a batch of test data $D_{test}^{K}$. However, the evolving system may result in the test data $D_{test}^{K}$ does not satisfy the independent and identically distribution with the training data, and the model $M^{K-1}$ trained using the initial training set $D^{ini}$ and $D_{train}^{K-1}$ may not be able to achieve effective detection of $D_{test}^{K}$. 
Meta-learning enables the effective generalization of evolving data. In recent research, MetaLog utilizes meta-learning for cross-system log anomaly detection tasks with stronger distribution shifts. Inspired by \cite{metalog},  we follow the classical MAML \cite{maml} architecture to implement the online meta-learning log anomaly detection model.

\textbf{Online Meta-tasks} Meta-learning performs gradient updating of the model by selecting samples from training and test data to compose the meta-task. Therefore, in the online updating framework based on the next event prediction, for each batch of test data, we have to select normal samples with the same distribution as the test data as training samples. First, we filter the test data by the normality detection model $N$ mentioned in Section \ref{sec:2.2} to obtain high-confidence normal samples as the online training samples for the test data. Second, we split each batch of data into different meta-task T, with the support set $D_{S}^{K, T}$ in each meta-training task originating from normal samples $D_{train}^{K}$, and the query set from $D_{Q}^{K, T}$. After determining the Query set $D_{Q}^{K, T}$, the support set $D_{S}^{K, T}$ is selected by adopting a temporal nearest neighbor strategy to ensure that the samples in the support set and the query set are similar. It is easy to understand that the more similar the training data is to the test data, the better the detection effect is. Therefore, we select the n samples in the $D_{train}^{K}$ that is temporally closest to the query set $D_{Q}^{K, T}$ as the support set $D_{S}^{K, T}$ to form an online meta-training task. 
It is worth noting that the online training data used are high-confidence normal sequences obtained from the normality detection model. Therefore, the number of support samples is not constrained by few-shot learning, but only by the number of normal samples in the test data.

\textbf{Online Meta-learning Detection}
The training process of meta-learning usually organizes multiple meta-tasks into one task batch for training. Therefore, we form a task batch of all the meta-tasks contained in the test data to fine-tune the detection model using meta-training. Unlike MAML, our work is to update the weights based on meta-training only in the online detection phase. To capture the global features of the log data, on the initial training data $D^{ini}$, we obtain the model $M^{ini}$ by standard training approach. For the K-th detection task, we fine-tune it to fit the local features employing meta-training. 

Specifically, in the initial training phase, instead of partitioning the data into support and query sets, we utilize the normal samples to train the normality detection model and the anomaly detection model separately. The normality detection model follows the approach outlined in~\cite{logonline}, whereby an auto-encoder algorithm is employed to enable online training through the acquisition of high-confidence normal samples in a non-manual manner. The anomaly detection model is implemented based on the LSTM algorithm, with the pre-training of the unsupervised detection model  $M^{ini}$ carried out using the next event prediction method described in Section~\ref{sec:2.2}. During the online meta-detection phase, as mentioned previously, we divide the test data of the K-th task batch into T online meta-tasks. We then fine-tune and detect the pre-trained anomaly detection model by meta-training to achieve rapid generalization of the model. In this process, each meta-task T corresponds to an episode, wherein we first use the support set $D_{S}^{K, T}$ to update the models $M^{K}$, and then input the query set $D_{Q}^{K, T}$ to obtain the detection results. After completing the episode on the T meta-tasks, the online learning and testing of the task batch is completed. Thus, for a task batch K, its meta-training loss can be defined as:
\begin{align}
\label{E4}
\mathcal{L}^{K}=\frac{1}{K}\cdot\frac{1}{T}\sum_{D_{Q}^{K,J}\in D_{Q}^{K}}\sum_{(X_{i},X_{i})\in D_{Q}^{K,J}}\left(M^{K,T}(X_{i})-Y_{i}\right)^{2}
\end{align}
Notably, the support and query set in each meta-task are matched using the temporal nearest neighbor strategy, which aligns with the application of the nearest neighbor log sample similarity phenomenon discussed in Section~\ref{sec:2.3}. Consequently, the distribution gap between the test data in the query set and the training data in the support set is minimized, enabling the model to achieve effective generalization to the evolving data.

\section{Experiment}
\label{sec:5}
In this section, we evaluate our approach by answering the following research questions:

RQ1: How effective is the OMLog method in stable systems?

RQ2: How effective is the OMLog method in evolving systems?

RQ3: How efficient is the OMLog method in different systems?

RQ4: How does each hyperparameter affect OMLog?

RQ5: How does each component contribute to OMLog?
\subsection{Experimental Setup}
\label{sec:5.1}
\subsubsection{Datasets}

\begin{table*}[ht]
\centering
\caption{Description of log datasets used in the experiments}
\begin{tabular}{cccccccccc}
\toprule  
Datasets&Parsing&Logs&Event&Time Span&	Grouping&	Window size&Step size&Types&Samples \\
\midrule
\multirow{2}*{HDFS}&\multirow{2}*{Public~\cite{hfaw}}&\multirow{2}*{11,175,629}&\multirow{2}*{28}&\multirow{2}*{38.7 hours}&\multirow{2}*{Session}&\multirow{2}*{/}&\multirow{2}*{/}&Normal&575,059 \\
\multirow{2}*{}&\multirow{2}*{}&\multirow{2}*{}&\multirow{2}*{}&\multirow{2}*{}&\multirow{2}*{}&\multirow{2}*{}\multirow{2}*{}&\multirow{2}*{}&Abnormal&16,838 (2.8$\%$) \\
\multirow{2}*{BGL}&\multirow{2}*{Drain~\cite{drain}}&\multirow{2}*{4,747,963}&\multirow{2}*{7,223}&\multirow{2}*{214.7 days}&\multirow{2}*{Sliding}&\multirow{2}*{100}&\multirow{2}*{100}&Normal&42,428 \\

\multirow{2}*{}&\multirow{2}*{}&\multirow{2}{*}{}&\multirow{2}{*}{}&\multirow{2}{*}{}&\multirow{2}{*}{}&\multirow{2}{*}{}&\multirow{2}*{}&Abnormal&4,745 (10.0$\%$) \\ 
\bottomrule
\label{table1}
\end{tabular}
\end{table*}

To evaluate the proposed OMLog, we use the stable dataset HDFS and the evolved dataset BGL as data sources, respectively. The details of each raw dataset are as follows:

The \textbf{HDFS dataset} \cite{pca} consists of 11,175,629 logs produced from more than 200 Amazon EC2 nodes spanning 38.7 hours. As shown in Table~\ref{table1}, these log messages form different log samples according to their block ID, including 575,059 normal log sequences and 16,838 abnormal log sequences (2.8$\%$). 

The \textbf{BGL Dataset} \cite{bgltbird} consists of 4,747,963 logs generated by a supercomputing system Blue-Gene/L spanning 214.7 days. Each message in the BGL dataset was manually labeled as either normal or anomalous. As shown in Table~\ref{table1}, we use the session window and sliding window to divide the log sequence respectively and label the log sequence as abnormal if there are any faulty logs in the sequence. Besides, we set the size of the sliding window to 100 and the step size to 100, and a total of 42,428 log sequence samples were obtained by this method, which contains 4,745 (10.0\%) anomalous sequences.
.

\subsubsection{Evaluation Metrics}
The essence of online log anomaly detection remains the binary classification of samples, and the test results are categorized into four types:

\textit{TP} is the number of abnormal samples that are correctly detected as abnormal samples. \textit{FN} is the number of abnormal samples that are not detected by the model. \textit{TN} is the number of normal samples that are correctly detected as normal samples. \textit{FP} is the number of normal samples that are wrongly identified as abnormal samples. 

To measure the effectiveness of models in log anomaly detection, we use Precision (PRE), Recall (REC), and F1-Score (F1) as detection metrics. Among them, precision is the proportion of all log sequences identified as anomalous that are correctly identified as anomalous. Precision = $\frac{TP}{TP+FP}$ Recall is the proportion of all log sequences that are correctly identified as anomalous. Recall = $\frac{TP}{TP+FN}$ F1-Score is the summed average of precision and Recall. F1-Score = $\frac{2*Precision*Recall}{Precision+Recall}$

\subsubsection{Implementations and Environments}
In our experiments, we process the log data and conduct log-based anomaly detection as follows: 

we use the open-source Spell\cite{spell} to implement log parsing, use SGD as the optimizer, set the learning rate to 0.00001, and set the epoch to 100. 
Our normality and anomaly detection models follow the work of \cite{logonline}, where the normality detection model was trained using an auto-encoder with a confidence threshold of 0.02. The anomaly detection algorithm model is a simple LSTM neural network.

Our approach is compared with supervised methods (e.g., LogRobust, CNN \cite{cnnlog}, ROEAD), unsupervised methods (e.g., DeepLog and LogAnomaly), and semi-supervised methods (PLELog and LogOnline) both online (e.g., ROEAD and LogOnline) and offline (e.g., LogRobust, CNN, DeepLog, LogAnomaly and PLELog). 

The supervised method ROEAD is performed as an online detection following 5 subtasks. For the LogOnline and the proposed OMLog methods, we set the epoch to 100. After every 20 training epochs, we conducted an online test and retained the best results, i.e. early stopping. We executed 5 independent experiments for each method, and the final experimental results are the average of the optimal outcomes from each experiment.

To complete the testing of the method, the experiments are performed using the following hardware and software platforms: Intel(R) Xeon(R) Gold 5218 CPU @ 2.30GHz, CPU cores 16, 256 GB RAM, NVIDIA Tesla V100-PCIE-32GB*2, CentOS Linux release 7.9, CUDA 11.7, cuDNN 7.6.5, Python 3.7.12, and PyTorch 1.7.0. All of the conducted experiments took about 1.5 months to complete, using two physical GPUs.

\subsection{RQ1: Detection Performance on Stable Log Data}
\label{sec:5.2}

\begin{table}[ht]
\newcommand{\tabincell}[2]{\begin{tabular}{@{}#1@{}}#2\end{tabular}}
\centering
\begin{threeparttable}
\caption{Results of the existing LAD approaches on HDFS dataset}
\begin{tabular}{ccccccc}
\toprule  
\multirow{2}*{Methods} & \multicolumn{3}{c}{train ratio = 0.5} & \multicolumn{3}{c}{train ratio = 0.8}\\
\cline{2-7}~ &PRE & REC & F1 & PRE & REC & F1\\
\midrule
LogRobust&	92.4&	99.7&	95.9&	93.4&	99.3&	96.3\\
CNN	&93.7&	99.9&	96.7&	94.4&	99.9&	97.1\\
ROEAD&	95.4&	99.3&	97.3&	97.3&	99.7&	98.5\\
\bottomrule
DeepLog&	81.7&	99.4&	89.7&	96.2&	86.6&	91.1\\
LogAnomaly&	86.7&	91.8&	89.2&	97.8&	85.7&	91.4\\
PLELog&	85.6&	98.2&	91.5&	97.7&	86.1&	91.5\\
LogOnline-1&	92.1&	88.0&	90.0&	94.9&	88.8&	91.7\\
LogOnline-Batch&	92.1&	88.0&	90.0&	94.9&	88.8&	91.7\\
\bottomrule
Ours&	95.6&	91.9&	93.7&	94.7&	92.8&	93.7\\
\bottomrule
\end{tabular}
\label{table2}
\end{threeparttable}
\end{table}

The HDFS dataset is derived from an unmodified Hadoop system and spans only 38.7 hours, with few log types and no evolution \cite{spine, logonline}. Therefore, we chose the original HDFS dataset as the data source for stabilized logs. For a fair comparison with existing log anomaly detection algorithms, we use 50\% and 80\% of the session sequences as the training set, and the remaining data as the test set to validate the precision, recall, and F1-Score of the different methods, respectively. 

As shown in Table \ref{table2}, the F1-Score of all kinds of methods on the HDFS dataset exceeds 90\%. Among them, the ROEAD method based on online supervised learning achieves optimal detection performance. The F1-Score of ROEAD reaches 98.5\% when the proportion of training data is 80\%. Moreover, the other two supervised learning methods, LogRobust and CNN, have F1-Score values of 96.3\% and 97.1\%, respectively, with recall rates above 99.0\%. This means that the supervised learning model recognizes essentially all anomalies on a stable dataset and maintains a very low level of false positives. Of course, this advantage relies on large and precious anomaly labels. The classical unsupervised anomaly detection methods DeepLog and LogAnomaly have the lowest detection performance. The F1-Score of both methods is below 90\% when the training set proportion is 50\%. Besides, the online method LogOnline slightly outperforms DeepLog and LogAnomaly in terms of detection performance. when the proportion of the training set is 80\%, the method is almost the same as the semi-supervised PLELog detection performance. 

 Our proposed OMLog method, trained using only normal samples, exceeds other unsupervised and semi-supervised detection methods, although it is lower than supervised learning methods in terms of detection effectiveness. The F1-Score of OMLog is 93.7\% for both when the training set proportion is 50\% and 80\%, respectively, outperforming the 90.0\% and 91.7\% of LogOnline. 

\subsection{RQ2: Detection Performance on Evolving Log Data}
\label{sec:5.3}

\begin{table}[ht]
\newcommand{\tabincell}[2]{\begin{tabular}{@{}#1@{}}#2\end{tabular}}
\centering
\begin{threeparttable}
\caption{Results of the existing LAD approaches on BGL dataset}
\begin{tabular}{ccccccc}
\toprule  
\multirow{2}*{Methods} & \multicolumn{3}{c}{train ratio = 0.5} & \multicolumn{3}{c}{train ratio = 0.8}\\
\cline{2-7}~ &PRE & REC & F1 & PRE & REC & F1\\
\midrule
LogRobust&	25.0&	48.0&	32.9&	37.4&	76.1&	50.2\\
CNN	&28.7&	49.3&	36.3&	55.3&	76.9&	64.3\\
ROEAD&	70.2	&47.8&	56.8	&74.2&	47.5&	57.9\\
\bottomrule
DeepLog	&13.7&	99.2&	24.1&	15.9&	100.0&	27.4\\
LogAnomaly&	15.7&	99.3&	27.1&	16.8&	100&	28.8\\
PLELog	&28.2	&54.7	&37.2&	40.9&	37&	38.9\\
LogOnline-1&	11.7&	98.7&	20.9&	16.1&	100.0&	56.7\\
LogOnline-Batch&	44.3&	78.2&	56.5&	46.5&	72.8&	27.8\\
\bottomrule
Ours	&56.3&	76.4&	64.8	&62.2&	77.2&	68.9\\
\bottomrule
\end{tabular}
\label{table3}
\end{threeparttable}
\end{table}

As demonstrated in Section \ref{sec:2.3}, we utilize the dataset BGL with significant distribution shift as the data source for the evolved system and set the training ratio to 50\% and 80\%, respectively.

As shown in Table \ref{table3}, when the training ratio is 80\%, the number of evolving log events in the test set is lower and the model learns the normal log sequence patterns more adequately. Therefore, the effect of evolved data on the detection method is easier to observe when the training ratio is 50\%. When the training ratio is 50\%, the detection performance of all offline detection methods lags behind that of online detection methods. Although the offline unsupervised methods DeepLog and LogAnomaly achieve a recall of more than 99.0\%, the extremely high false positives prevent their detection performance from being adopted, with F1-Score below 40\%. Although the offline semi-supervised and supervised methods PLELog, LogRobust, and CNN outperform the unsupervised detection methods, the F1-Scores are still below 40\%. 
Online detection can adapt to emerging log sequence patterns through a continuous model update mechanism. Compared to online detection methods, LogOnline, ROEAD, and our proposed OMLog exhibit higher accuracy and F1-Score when dealing with evolving log data. It is easy to observe that on the log dataset BGL, utilizing 50\% and 80\% of the data for training, OMLog achieves the F1-Score of 64.8\% and 68.9\%, respectively, which significantly outperforms the SOTA online semi-supervised detection method, LogOnline. Even when compared to ROEAD, a supervised method that utilizes sufficient labels for training, OMLog still maintains its leading performance. 

Notably, OMLog is based on online meta-learning, and its detection performance relies more on local information. Compared to LogOnline, which is also based on online learning and next-event prediction, OMLog is better at generalizing to novel event features and achieving improved detection results.

\subsection{RQ3: Detection Efficiency of Different Approaches}
\label{sec:5.4}

\begin{table}[ht]
\newcommand{\tabincell}[2]{\begin{tabular}{@{}#1@{}}#2\end{tabular}}
\centering
\begin{threeparttable}
\caption{Comparison of training and testing times for different methods on HDFS Dataset}
\begin{tabular}{ccccc}
\toprule  
\multirow{2}*{Methods} & \multicolumn{2}{c}{train ratio = 0.5} & \multicolumn{2}{c}{train ratio = 0.8}\\
\cline{2-5}~&train(m) & test(s) &  train(m) & test(s) \\
\midrule
LogRobust&	7.7&	4.3	&8.7&3.6\\
CNN	&14.7&	38.4&	27.5&	17.6\\
ROEAD&	44.5&	10.8&	54.1&	7.3\\
DeepLog&	120.5&	151&	182.8&	103\\
LogAnomaly&	285&	351&	465&	309\\
PLELog	&41.5&	4.1&	71.5	&3.2\\
LogOnline-1 &119.2&	11,576&	152.3&	8,972\\
LogOnline-Batch	&123.3	&259	&156.5	&218\\
\bottomrule
Ours&	124.7&	148&	150.2&	122\\
\bottomrule
\end{tabular}
\label{table4}
\end{threeparttable}
\end{table}

\begin{table}[ht]
\newcommand{\tabincell}[2]{\begin{tabular}{@{}#1@{}}#2\end{tabular}}
\centering
\begin{threeparttable}
\caption{Comparison of training and testing times for different methods on BGL Dataset}
\begin{tabular}{ccccc}
\toprule  
\multirow{2}*{Methods} & \multicolumn{2}{c}{train ratio = 0.5} & \multicolumn{2}{c}{train ratio = 0.8}\\
\cline{2-5}~&train(m) & test(s) &  train(m) & test(s) \\
\midrule
LogRobust&	4.6&	1.1&	5.4&	0.27\\
CNN&	8.5&	5.9	&14.3&	2.4\\
ROEAD&	25.9&	2.1&	30.7&	0.45\\
DeepLog&	126.7&	103	&130.2&	94.2\\
LogAnomaly&	1,500&	830&	2,193.3&	635\\
PLELog	&14.4	&0.8	&2.1&	0.18\\
LogOnline-1 & 119.2&	9,496&	139.3&	8,064\\
LogOnline-Batch&	121.7&	234&	135&	186\\
\bottomrule
Ours&	123.3&	158&	132.5&	118.4\\
\bottomrule
\end{tabular}
\label{table6}
\end{threeparttable}
\end{table}

\begin{figure*}[!htb]
	\centering
	\scriptsize
	\subfloat[$\epsilon$-Performance,Train Ratio=0.5]{
		\includegraphics[width=0.24\linewidth]{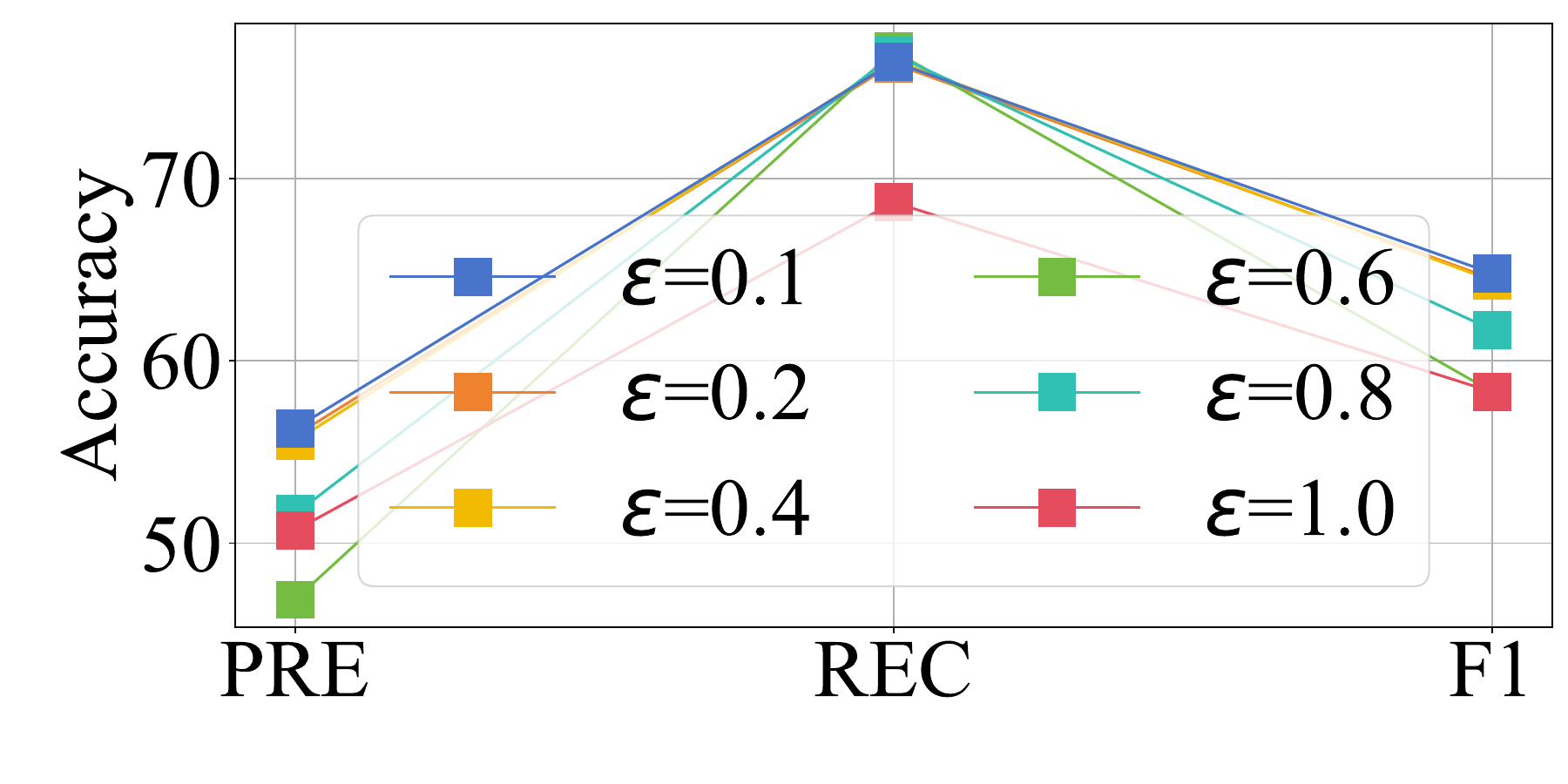}
	}
	\hfill
	\subfloat[$\epsilon$-Efficiency, Train Ratio=0.5]{
		\includegraphics[width=0.24\linewidth]{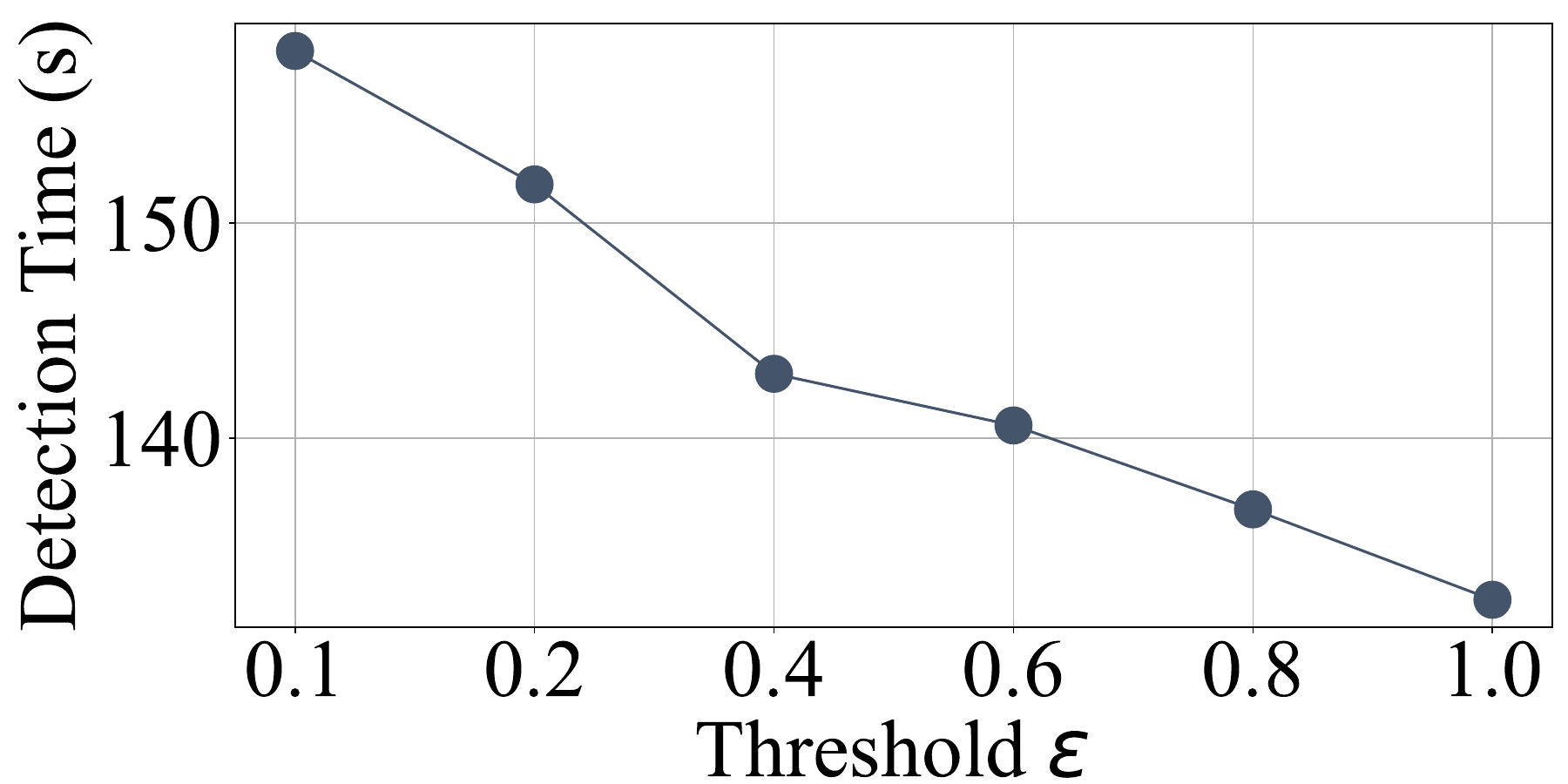}
	}
	\hfill
	\subfloat[$\epsilon$-Performance, Train Ratio=0.8]{
		\includegraphics[width=0.24\linewidth]{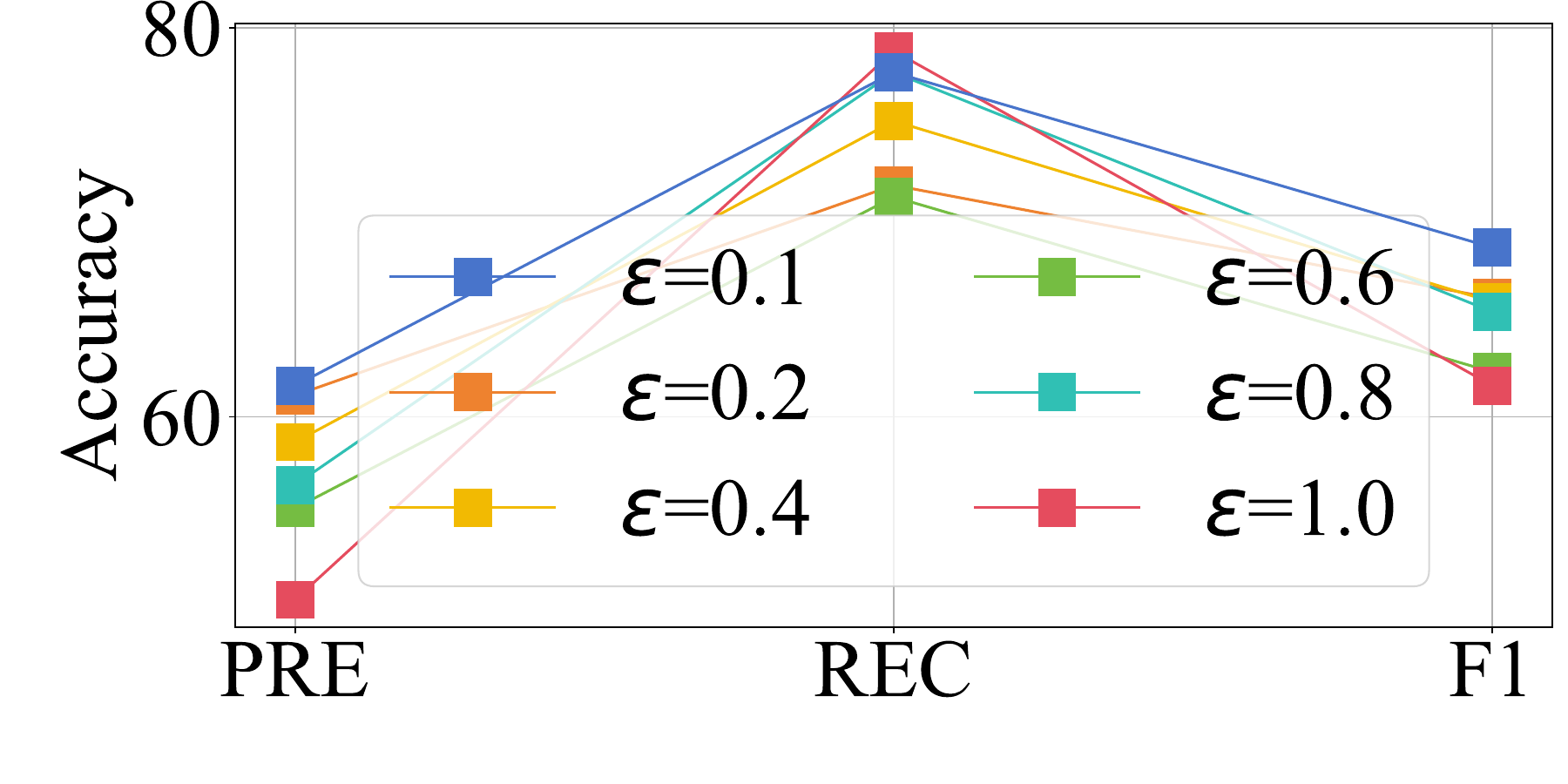}
	}
	\hfill
	\subfloat[$\epsilon$-Efficiency, Train Ratio=0.8]{
		\includegraphics[width=0.24\linewidth]{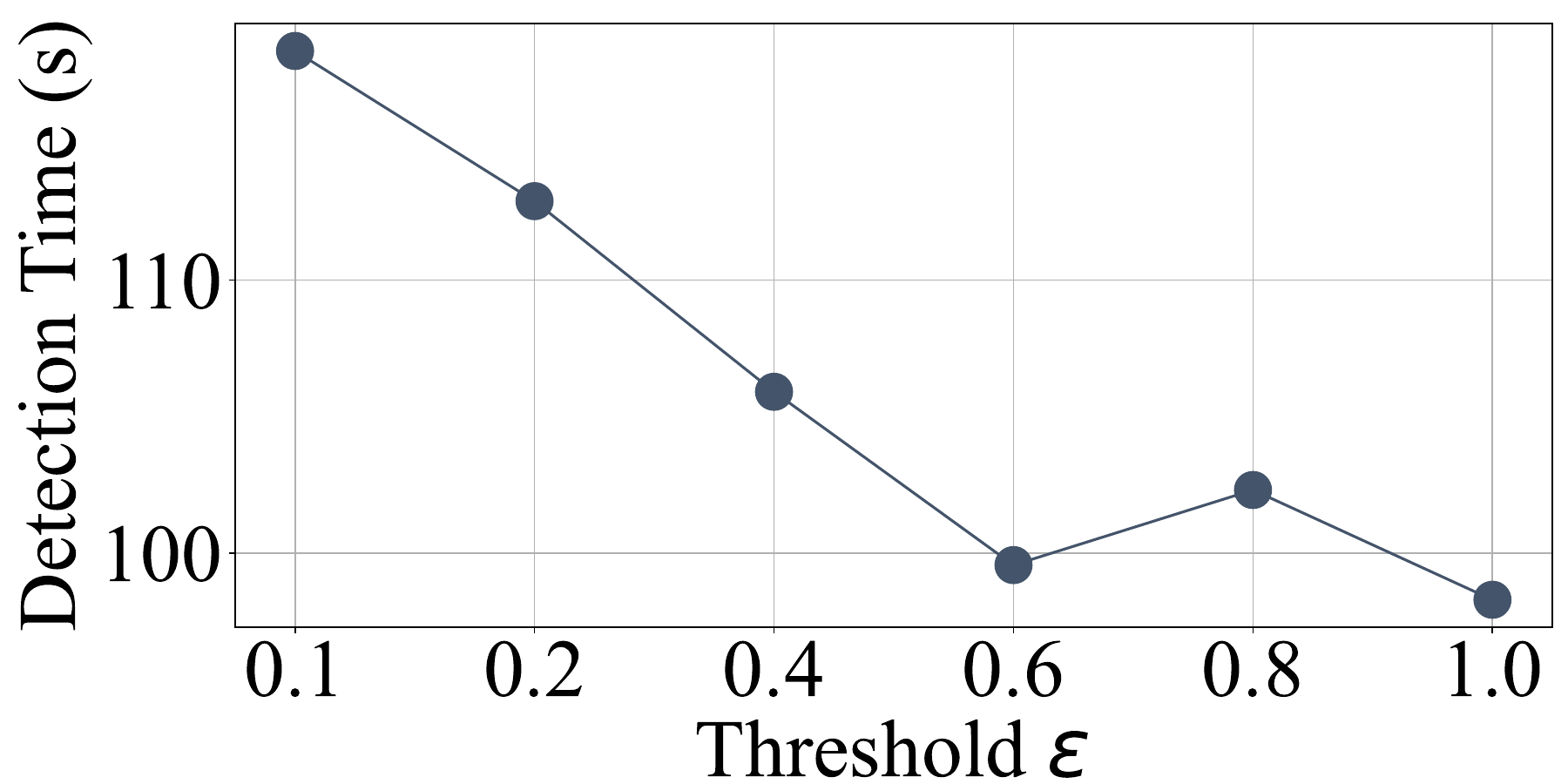}
	}

	\centering
	\scriptsize
	\subfloat[T-Performance, Train Ratio=0.5]{
		\includegraphics[width=0.24\linewidth]{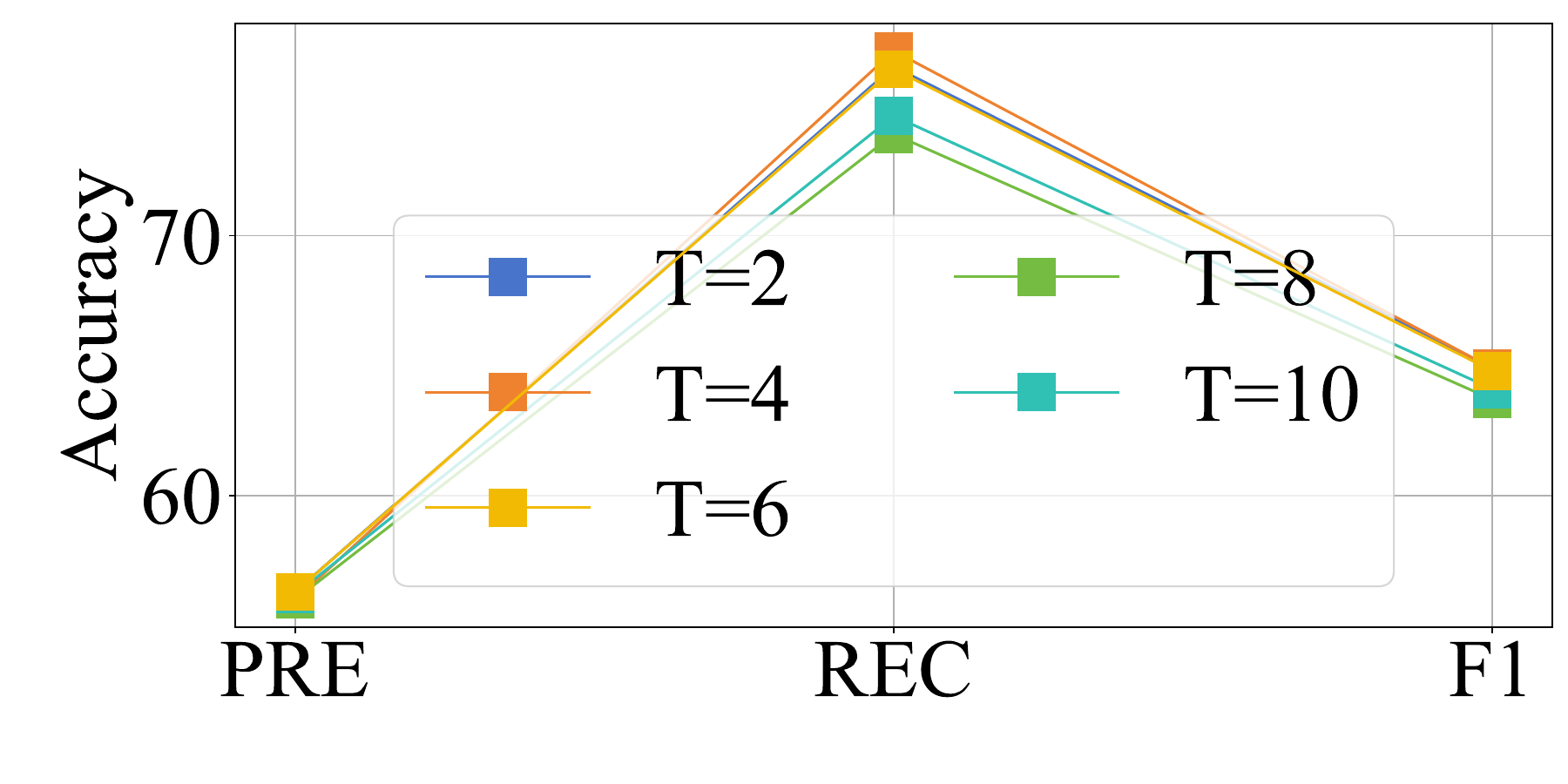}
	}
	\hfill
	\subfloat[T-Efficiency, Train Ratio=0.5]{
		\includegraphics[width=0.24\linewidth]{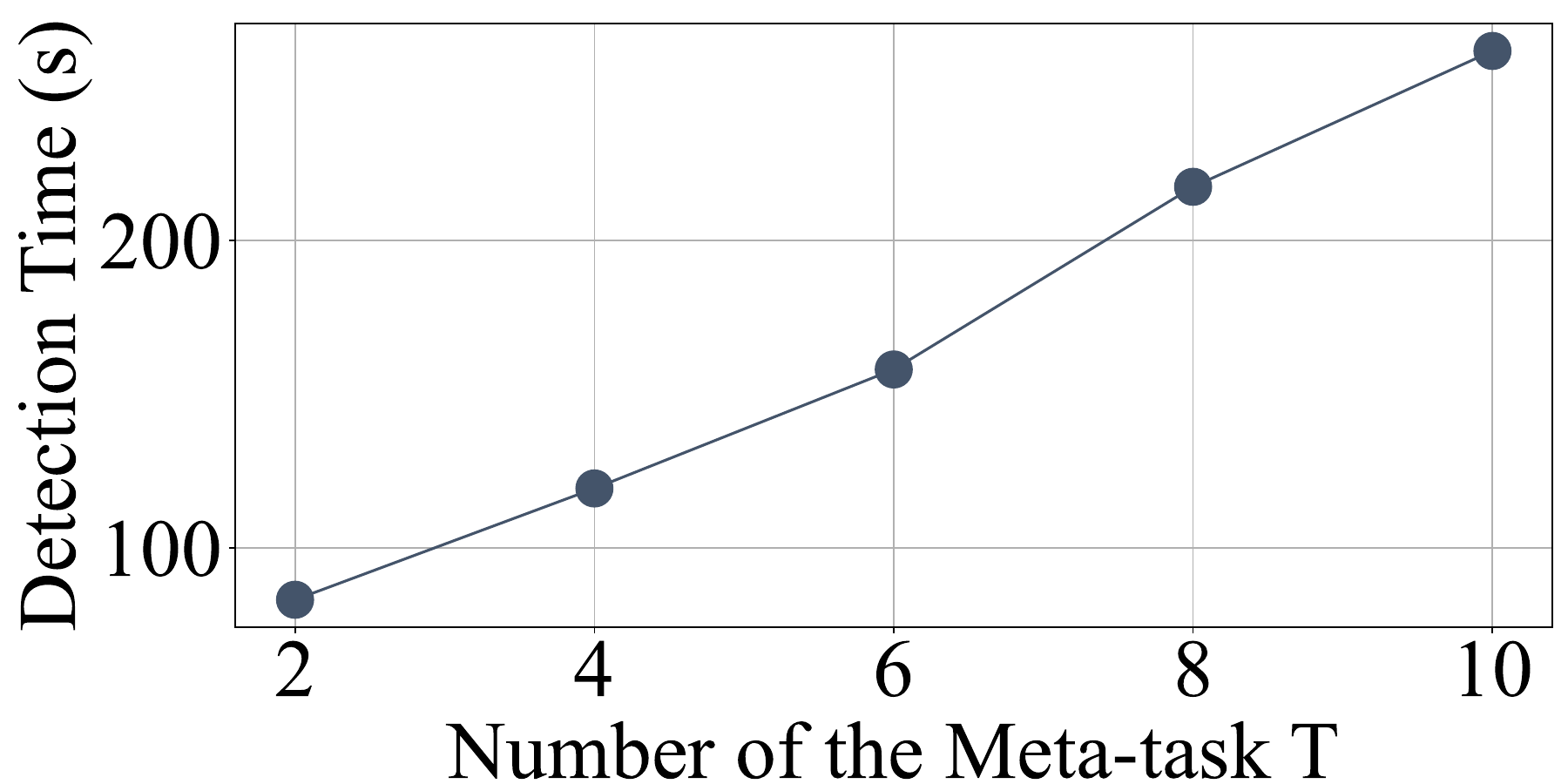}
	}
	\hfill
	\subfloat[T-Performance, Train Ratio=0.8]{
		\includegraphics[width=0.24\linewidth]{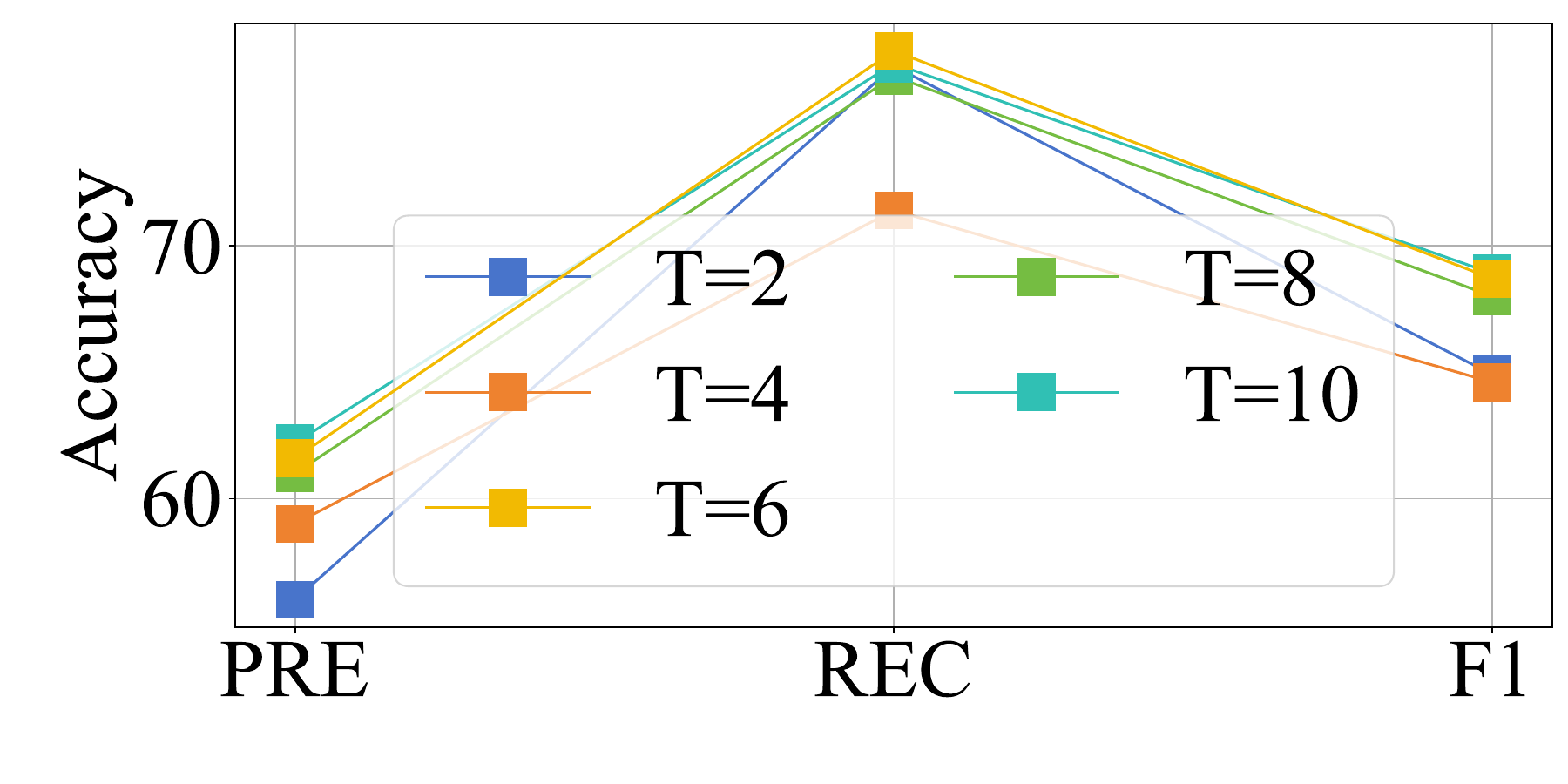}
	}
	\hfill
	\subfloat[T-Efficiency, Train Ratio=0.8]{
		\includegraphics[width=0.24\linewidth]{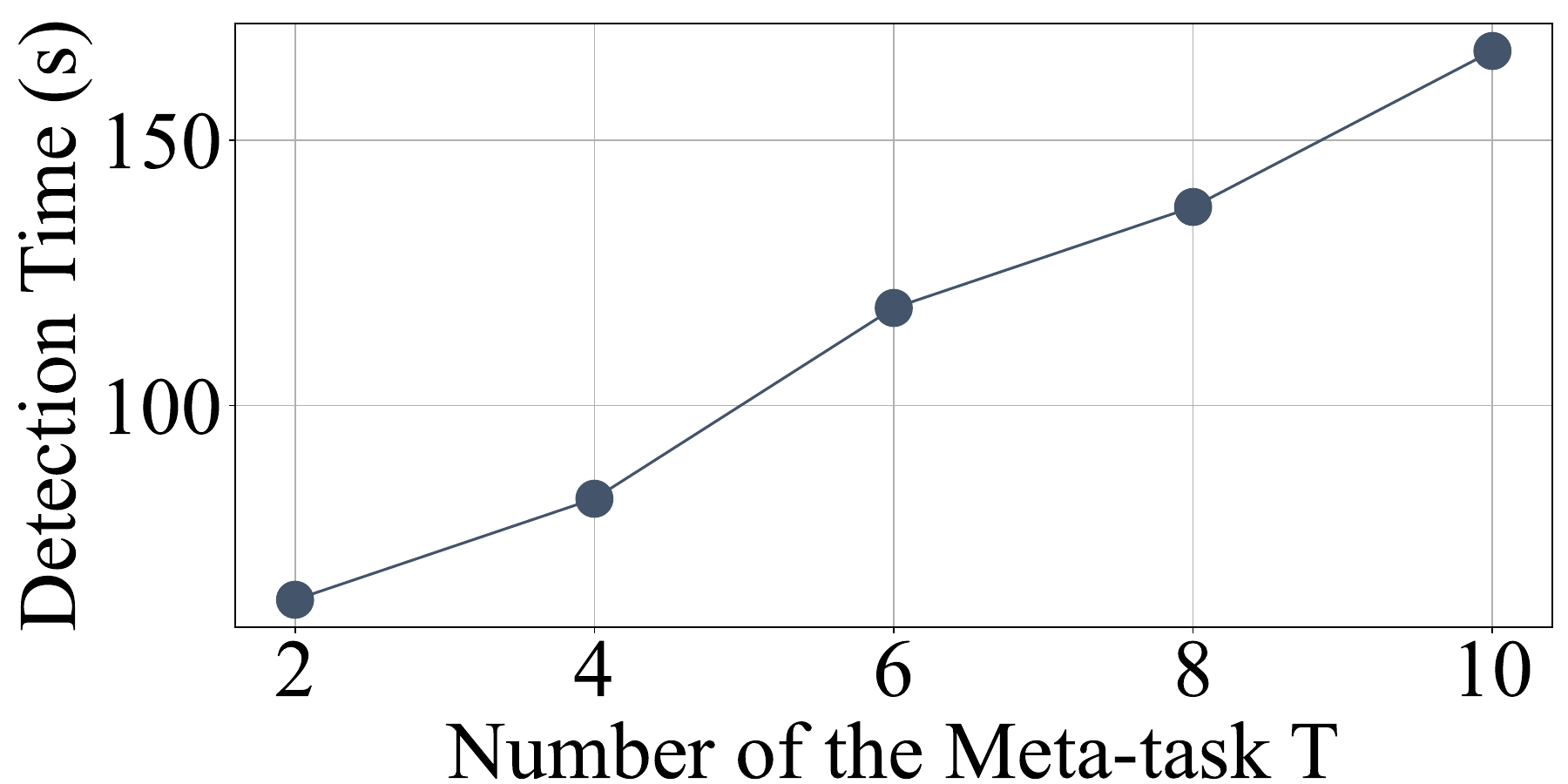}
	}

\caption{Impact of hyperparameters on detection performance and detection efficiency. Two hyperparameters were mainly investigated, including the DSD threshold $\epsilon$ and the number of meta-tasks in a Batch (T)}
\label{figure6}
\end{figure*}

Table \ref{table4} gives the training time and detection time of OMLog on HDFS and BGL, respectively. Overall, offline methods PLELog, LogRobust, CNN, and ROEAD are equally efficient with very short detection times. For example, LogRobust takes only 0.27s to detect 20\% of the data in the BGL dataset. Among the online detection methods, the supervised learning method ROEAD takes no more than 1 second to detect 20\% of the samples in the BGL dataset. However, the detection time of ROEAD does not include the time to manually label the system state online. Manual labeling is precious and time-consuming. Therefore the method is difficult to apply in practice.

We evaluate the detection efficiency of the semi-supervised method LogOnline by following the standard online learning mechanism (i.e. LogOnline-1, batch size = 1) and batch learning mechanism (i.e. LogOnline-Batch), respectively. It is easy to observe that LogOnline takes more than 5 hours to detect 20\% of BGL samples if the standard online learning mechanism is followed. If the method is based on batch learning for detection, the time consumed is greatly reduced to 186 seconds. Therefore, our direction of distribution shift detection and online meta-learning based on Batch data is reasonable. Compared with LogOnline, our proposed OMLog method does not rely on manual labeling and shows great potential in detection performance while the detection efficiency is much higher than standard online learning methods.

\subsection{RQ4: Detection Performance and Efficiency with Different Hyperparameters}
\label{sec:5.5}

We explored the impact of changes in the two core hyperparameters within the OMLog, the DSD threshold e and the number of meta-tasks, on the detection performance and efficiency for two different training ratio configurations of 0.5 and 0.8 on the BGL dataset. Figure \ref{figure6} shows a visual presentation of the experimental results.
The analysis shows a significant negative correlation between DSD threshold and detection performance, while it is positively correlated with detection efficiency. Specifically, in the 50\% training rate scenario, F1 decreases from 64.8\% to 58.3\% as the threshold increases from 0.1 to 1.0. Conversely, along with the increase of the threshold, the detection time decreases from 158 s to 132.5 s. Easy to understand, setting the DSD threshold too high means that for test data with distributional shifts from the training data, generalization cannot be achieved by the online learning mechanism, which may lead to a degradation of the detection performance. On the contrary, setting the DSD threshold too low means that the model carries out frequent online updates for test data without distributional shifts, which may lead to a decrease in detection efficiency.

On the other hand, the effect of the change in the number of meta-tasks on the detection performance is more complex, while it shows a direct negative correlation on the detection efficiency. For example, at 80\% training ratio, as the number of meta-tasks increases from 2 to 10, the F1 score improves from 64.9 to 68.9, which implies that increasing the number of meta-tasks helps to enhance the generalization ability of the model to the test data through more diverse support samples. However, it is interesting to note that the increase in the number of tasks does not result in a sustained performance improvement at 50\% training rate, but rather a performance degradation. Meanwhile, the detection time increased significantly, climbing from 63.4 to 166.8 seconds, implying that the higher the number of meta-tasks, the lower the detection efficiency. 
To summarize, reasonable adjustment of the DSD threshold and the number of meta-tasks is the key to balancing detection performance and efficiency. In practical applications, meticulous parameter tuning is required according to specific task requirements.

\subsection{RQ5: Contribution of Main Components in OMLog}
\label{sec:5.6}

\begin{table}[ht]
\newcommand{\tabincell}[2]{\begin{tabular}{@{}#1@{}}#2\end{tabular}}
\centering
\begin{threeparttable}
\caption{Results of ablation experiments}
\begin{tabular}{ccccccc}
\toprule  
\multirow{2}*{Methods} & \multicolumn{3}{c}{train ratio = 0.5} & \multicolumn{3}{c}{train ratio = 0.8}\\
\cline{2-7}~ &PRE & REC & F1 & PRE & REC & F1\\
\midrule
Offline	&18.1&	79.4&	29.6&	15.8&	99.1&	27.3\\
Online	&42.7&	76.8&	54.9&	51.6&	68.9&	59.0\\
Online+DSD	&40.4&	80.0&	53.6&	48.4&	77.5&	59.4\\
Meta-learning&	43.9&	76.6&	55.8&	53.7&	74.6&	62.4\\
OMLog	&56.3&	76.4&	64.8	&62.2&	77.2&	68.9\\
\bottomrule
\end{tabular}
\label{table5}
\end{threeparttable}
\end{table}

OMLog contains two main components, distribution shift detection (DSD) and online meta-detection (OMD). To investigate the impact of each component on the detection performance, we designed four different comparison experiments: 
\begin{itemize}
\item Offline, i.e., without the online update strategy;

\item Online, i.e., online updating using high-confidence samples from a batch of test data;

\item Online+DSD, i.e., combining DSD with online updating;

\item Meta-learning, i.e., using the MAML framework on both the initial training data and the test data.
\end{itemize}

As shown in Table \ref{table5}, the offline detection is completely unable to handle the system evolution effectively. The F1-Score of the offline method is below 30\% on different percentages of the training set. The F1-Score of online learning achieved 59.0\%, and the detection performance far exceeds that of offline learning, which proves the great potential of the online detection mechanism when facing the evolving system. The performance of the online detection method is essentially unchanged after adding the DSD component. This demonstrates that distribution shift detection does not affect the detection accuracy while detecting efficiency. OMLog makes a reasonable application of the observed local log event stability. Furthermore, in the comparison experiments with meta-learning, it can be observed that the detection accuracy of the OMLog method significantly exceeds that of the meta-learning method of MAML, which proves the importance of the combination of global and local information mentioned in our method. The OMLog method can quickly generalize to the evolutionary data while taking full advantage of the global information to ensure that the detection model can be discriminable.

\section{Discussion}
\label{sec:6}
\subsection{Why does OMLog Work?}
\label{sec:6.1}
There are two main reasons why OMLog performs better than state-of-the-art approaches. First, noting that local log events are stable, we perform distribution shift detection on batches of test data to determine whether to perform a time-consuming online learning strategy or fast offline inference. Second, noting that the near-neighbor log samples are similar, OMLog combines online learning with meta-learning. In the initial training phase, OMLog utilizes a large number of normal samples for standard training to capture global features. In the online detection phase, OMLog exploits the highly overlapping features of the near-neighbor log samples to construct meta-tasks for weight updating, so that the model can be effectively generalized to the evolving data.

Our study demonstrates the effectiveness of OMLog in evolving systems. However, OMLog still has limitations. Our approach trains and detects models on log data based on the next event prediction. The detection process needs to further determine whether the predicted event appears among the Top-K candidate events based on the probability distribution of the model output. However, this process is very time-consuming. In future work, we will consider more practical online detection frameworks to break the bottleneck of detection efficiency.
\subsection{Threats to Validity}
\label{sec:6.2}
We have identified the following major threats to validity.

\textbf{Data Quality:} Although the HDFS and BGL datasets have been widely used in previous studies and are publicly available, they were collected and introduced before 2007, which means that they may not contain all the relevant features and patterns associated with the current evolution of system logs. Therefore, drawing conclusions based on these datasets alone may introduce bias and does not fully guarantee their generalizability to real-world environments.

\textbf{Practicality:} While our OMLog approach shows significant improvements in performance and efficiency compared to other methods, it may not be comprehensive enough to address log anomaly detection in real software systems independently. Despite obtaining the F1-Score of 93.7\% and 64.9\% on the HDFS and BGL datasets, respectively, a certain number of false alarms still occur. Therefore, in practical applications, it may be necessary to use additional tools \cite{hilog} or incorporate human feedback to enhance the effectiveness of OMLog.

\section{Related Work}
\label{sec:7}
\subsection{Log Anomaly Detection}
Log anomaly detection is an important part of complex system maintenance. Detection methods for unstable evolving logs have received extensive attention, including semantic similarity-based methods \cite{plelog,swisslog,neurallog, hilbertlog, hitanomaly, spikelog,fslogtnsm, metalog} and dynamic update-based methods \cite{ada, roead, logonline, logflash, evlog}.


Semantic similarity-based LADs, which utilize semantic similarity to fit evolved log events, contain unsupervised (e.g., LogAnomaly \cite{loganomaly} and DeepSyslog \cite{deepsyslog}), supervised (e.g., LogRobust \cite{logrobust} and LightLog \cite{lightlog}), and semi-supervised methods (e.g., PLELog \cite{plelog} and SwissLog \cite{swisslog}). Noting that the One-Hot encoding of classical DeepLog does not consider the semantic similarity among logs, LogAnomaly employs word2vec to transform logs into semantic vectors, adding quantitative vectors to enrich features. DeepSyslog adapts to the log evolution by character-level word embedding to mine the hidden semantics and contextual information. LogRobust fuses TF-IDF  and FastText \cite{fasttext} to construct log semantic similarity vectors to reduce the impact of evolutionary and parsing noise, LightLog extends LogRobust by dimensionality reduction of semantic vectors, SwissLog combines semantic and temporal embeddings to capture anomalies and adapt to changes in the log format, and PLELog uses probabilistic markers to estimate soft labels of log sequences and reduces the impact of unstable data with semantic embeddings. stable data effects. Dynamic update-based LAD methods adapt to log evolution through incremental or online training, including unsupervised ADA \cite{ada}, supervised ROEAD \cite{roead}, and semi-supervised LogOnline \cite{logonline}. ADA uses adaptive model selection and dynamic thresholding for unsupervised detection and utilizes a Pareto strategy to select the optimal model threshold and update it periodically. ROEAD uses a robust feature extractor to remove noise and a support vector machine to update the model online. LogOnline uses the two-model strategy to train the detection model online using high-confidence normal logs captured by the normal detection model. 

However, on the one hand, semantic similarity-based detection methods cannot deal with semantically irrelevant log event types and frequently changing sequence patterns; on the other hand, online dynamic updating methods are less efficient in detection and difficult to generalize to evolving data effectively.

\subsection{Meta-learning}
Meta-learning is often understood as “learning to learn”, where a meta-model is pre-trained and quickly updated as new tasks come along to make the model perform better on unseen data \cite{mlsurvey1, mlsurvey2, mlsurvey3}. Current works close to our topic include two types of meta-learning tasks, online meta-learning and time-series based meta-learning. Online meta-learning proposes to integrate meta-learning with online learning. Models are expected to perform meta-learning in the order of tasks \cite{ometa,meoml}. Current research has proposed online gradient descent and following the regularized leader methods to address the online meta-learning problem to quickly adapt to all previously viewed task instances based on taking several gradient steps from the meta-model. Time series-based meta-learning has gained attention in recent years, DeepTime \cite{deeptime} presents an example of a fast and effective meta-optimization framework that augments a deep time-indexed model with a connected Fourier feature module to efficiently learn high-frequency patterns in a time series. The MetaLog \cite{metalog} approach introduced meta-learning for the first time to the field of log anomaly detection using a small number of log time sequence samples for cross-system detection. 

Encouraged by the above studies, we propose a semi-supervised detection framework based on the combination of online learning and meta-learning to capture global and local features of log samples for fast generalization while ensuring discriminability. 

\section{Conclusion}
\label{sec:8}
In this paper, we propose a semi-supervised online log anomaly detection method, OMLog. By designing a distribution shift detection algorithm based on maximum mean discrepancy and an online learning mechanism combining meta-learning, the method effectively solves the challenges posed by changes in the types and frequencies of log events in complex systems and exhibits a strong generalization capability to evolving data. We conducted extensive experiments on two public log datasets, the results demonstrate that OMLog, trained solely on normal log sequences, exhibits significant advantages in detection efficiency and outperforms existing SOTA log anomaly methods. In future work, we aim to explore more efficient detection frameworks to overcome existing limitations and improve the real-time processing capability of the model.

\bibliography{omlog.bib}
\bibliographystyle{IEEEtran}

\begin{IEEEbiography}[{\includegraphics[width=1in,height=1.25in,clip,keepaspectratio]{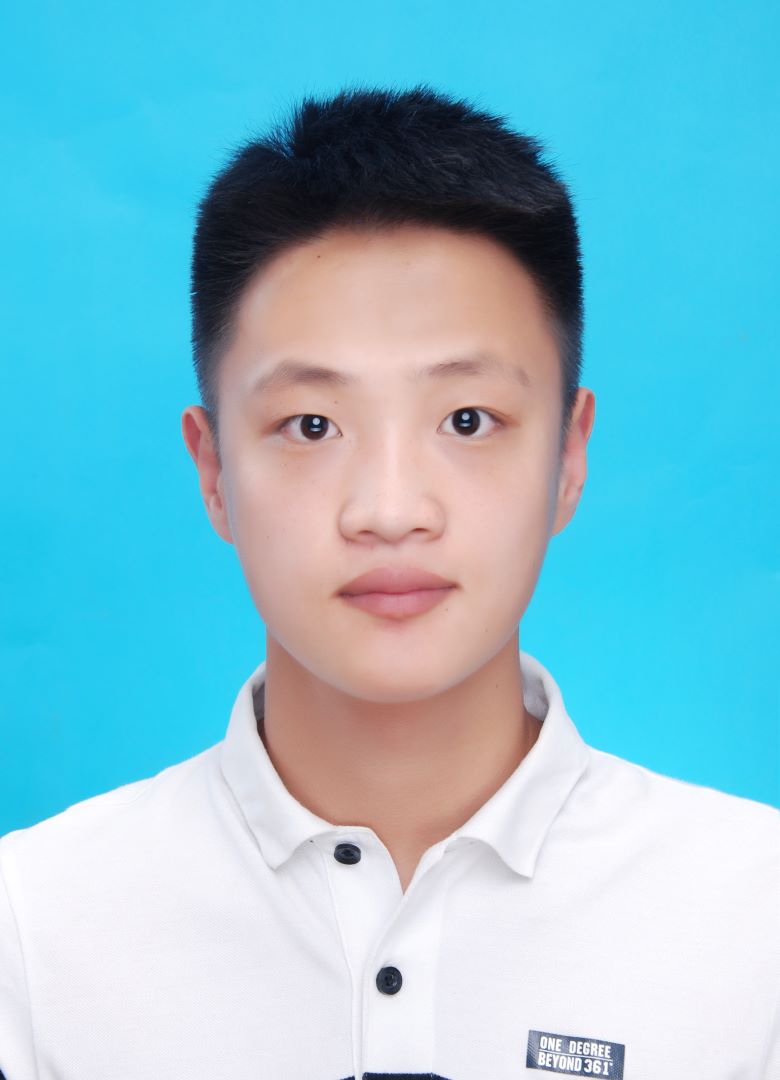}}]{Jiyu Tian} received the master's degree in cyberspace security from Dalian University, Dalian, China. He is currently pursuing the Ph.D. degree in Dalian University of Technology, Dalian, China. His current research focuses on log anomaly detection and cyberspace security.
\end{IEEEbiography}

\begin{IEEEbiography}
[{\includegraphics[width=0.9in,height=1.25in,clip,keepaspectratio]{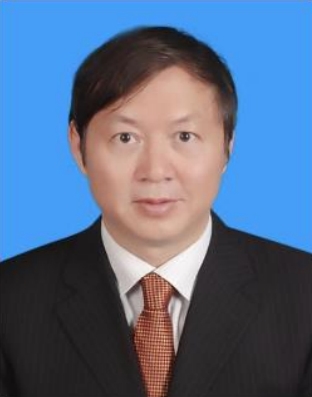}}]{Mingchu Li} received his doctorate in Mathematics, University of Toronto in 1998. He worked for School of Software of Tianjin University as a full professor (from 2002 to 2004), for School of Software Technology of Dalian University of Technology as a full Professor and Vice Dean (from 2002 to 2023), for Jiangxi Normal University as a full Professor from 2024 to now.  His main research interests include theoretical computer science and information security, and trust models and cooperative game theory.
\end{IEEEbiography}

\begin{IEEEbiography}
[{\includegraphics[width=1in,height=1.25in,clip,keepaspectratio]{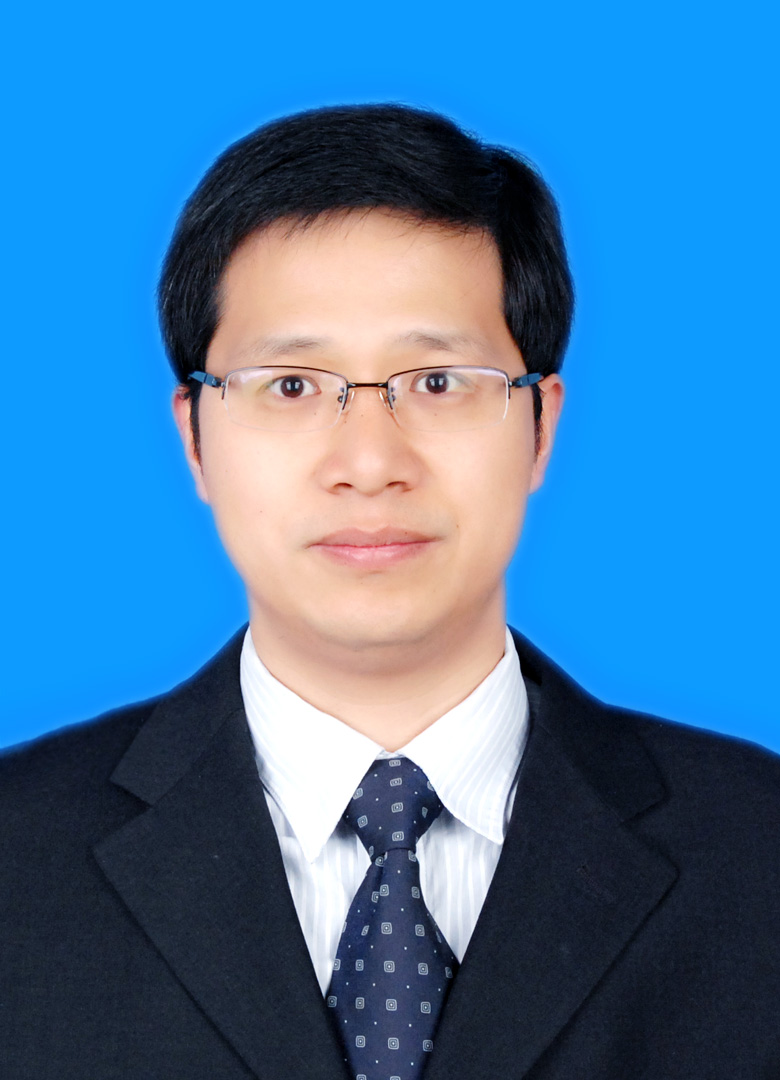}}]{Zumin Wang} is a professor at Dalian University since 2014. His research interests include Sensors and Sensor Application, Wireless Sensor Networks, and Smart City. He received his MS degree in Mechanical Manufacturing and Automation in 2004 from North University of China, and Ph.D. degree in Physical Electronics in 2007 from the Chinese Academy of Sciences. He was a visiting scholar at the University of Washington from Nov. 2016 to Nov. 2017. He is a distinguished member of CCF.
\end{IEEEbiography}

\begin{IEEEbiography}
[{\includegraphics[width=1in,height=1.25in,clip,keepaspectratio]{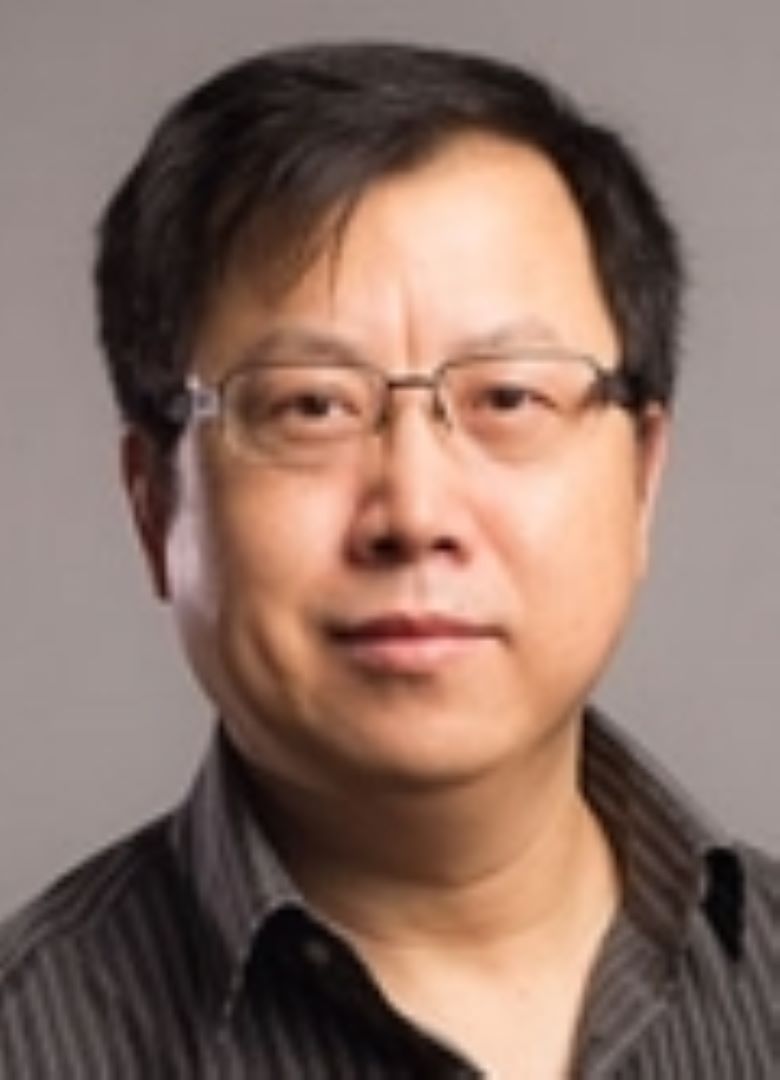}}]{Liming Chen} received his B.Eng and M.Eng degrees from Beijing Institute of Technology, China, in 1985 and 1988 respectively, and his Ph.D. degree from De Montfort University, UK, in 2003. He is currently a Chair Professor at the School of Computer Science and Technology, Dalian University of Technology, China. His research interests include pervasive computing, data analytics, artificial intelligence, user-centred intelligent cyber-physical systems and their applications in smart healthcare and cyber security. He has over 300 publications in the aforementioned areas. Liming is an IET Fellow and a Senior Member of IEEE.
\end{IEEEbiography}

\begin{IEEEbiography}
[{\includegraphics[width=1in,height=1.25in,clip,keepaspectratio]{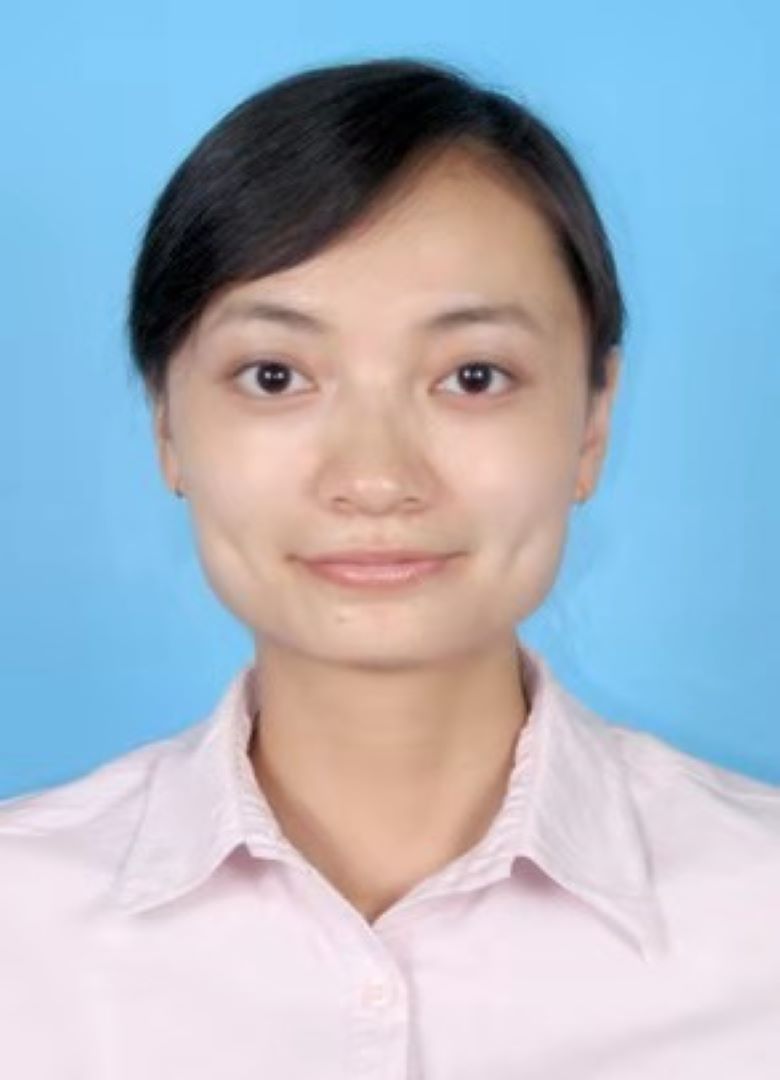}}]{Jing Qin} received the Ph.D degree in the School of Computer Science and Technology, Dalian University of Technology, Dalian, China. She is an associate professor in the School of Software Engineering, Dalian University. Her research interests include multimedia information retrieval, signal processing and machine learning.
\end{IEEEbiography}

\begin{IEEEbiography}
[{\includegraphics[width=1in,height=1.25in,clip,keepaspectratio]{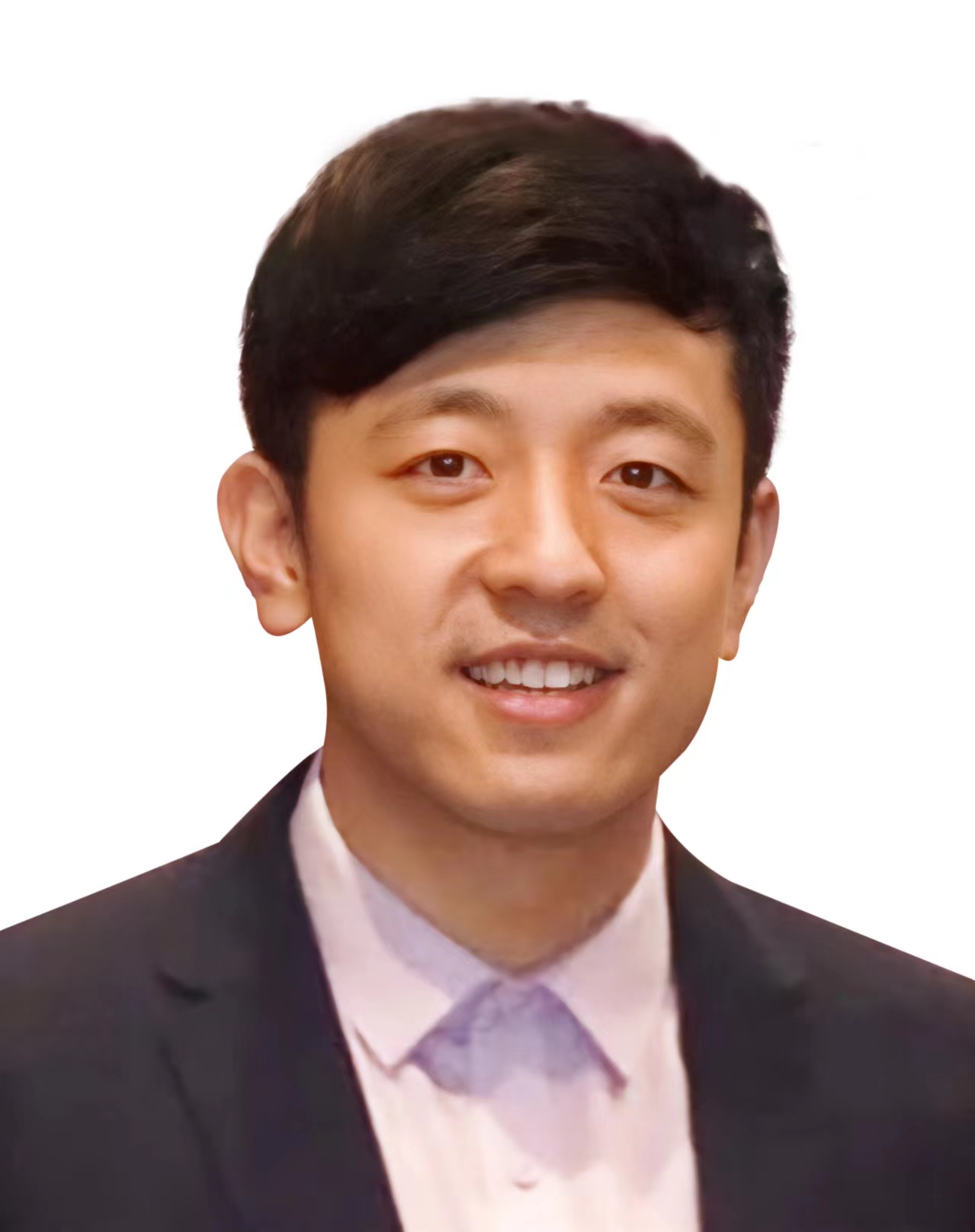}}]{Runfa Zhang} 
Runfa Zhang received the B.S. degree in mathematics from Taiyuan Normal University, in 2015. He received his M.S. degree in mathematics from Inner Mongolia University of Technology, in 2020 and his Ph.D. degree in software from Dalian University of Technology, in 2024. He worked for the School of Automation and Software Engineering, Shanxi University as an Associate professor from 2024 to now. His main research interests include neural network, symbolic computation, nonlinear waves, cyber security, and game theory.
\end{IEEEbiography}

\vfill
\end{document}